\documentclass{elsarticle}
\makeatletter
\def\ps@pprintTitle{%
 \let\@oddhead\@empty
 \let\@evenhead\@empty
 \def\@oddfoot{\centerline{\thepage}}%
 \let\@evenfoot\@oddfoot}
\makeatother
\usepackage{lineno,hyperref}

\biboptions{authoryear}

\usepackage{graphicx}
\usepackage{amsbsy}
\usepackage{amsmath}
\usepackage{amssymb}
\usepackage{amsfonts}
\usepackage{natbib}
\usepackage{amssymb}
\usepackage{array}
\usepackage[usenames,dvipsnames]{xcolor}

\usepackage{fullpage}

\newcommand{\bbeta}{\boldsymbol{\beta}}
\newcommand{\by}{\boldsymbol{y}}
\newcommand{\tr}{\mathrm{tr}}
\newcommand{\bx}{\boldsymbol{x}}

\begin{document}

\begin{frontmatter}

\title{Model selection via Bayesian information capacity designs for generalised linear models}



\author[UoS]{David C. Woods\corref{correspondingauthor}}
\cortext[correspondingauthor]{Correspondence to: Mathematical Sciences, University of Southampton, Southampton, SO17 1BJ, UK.}
\ead{D.Woods@southampton.ac.uk}
\author[QUT]{James M. McGree}
\author[UoS]{Susan M. Lewis}
\address[UoS]{University of Southampton, UK}
\address[QUT]{Queensland University of Technology, Australia}

\begin{abstract}
The first investigation is made of designs for screening experiments where the response variable is approximated by a generalised linear model. A Bayesian information capacity criterion is defined for the selection of designs that are robust to the form of the linear predictor. For binomial data and logistic regression, the effectiveness of these designs for screening is assessed through simulation studies using all-subsets regression and model selection via maximum penalised likelihood and a generalised information criterion. For Poisson data and log-linear regression, similar assessments are made using maximum likelihood and the Akaike information criterion for minimally-supported designs that are constructed analytically. The results show that effective screening, that is, high power with moderate type I error rate and false discovery rate, can be achieved through suitable choices for the number of design support points and experiment size. Logistic regression is shown to present a more challenging problem than log-linear regression. Some areas for future work are also indicated.\\[1ex]
This paper will appear in Computational Statistics and Data Analysis. 
\end{abstract}


\end{frontmatter}


\section{Introduction}\label{sec:intro}








An important problem in scientific discovery is to find those variables (or factors) that have a substantive influence on an observed response through experiments on a possibly large set of potentially important variables. There has been much research into such variable screening, or model selection, focussed on the design and analysis of experiments in which the response variable is adequately approximated by a linear model (see \citealp{DWDLV2014} and \citealp{WL2016}, and references therein). Such experiments are used increasingly in scientific research and product development, for example, in the pharmaceutical and chemical industries.

In many practical applications, for example when binary or count data are observed, a generalised linear model (GLM; \citealp{MN1989}) may be needed to describe a response. Previous research on designs for model selection for GLMs has focussed on experiments involving only a few variables through pairwise comparisons of a small number of models (see, for example, \citealp{LTT2007} and \citealp{WWEL2008}). Hence, such methods are not applicable to, or easily generalisable for, the screening problem. In the literature, the majority of multi-variable experimentation with GLMs has employed (fractional) factorial designs, including examples on solder-joint defects \citep{HN1997}, windshield molding, non-conforming tiles and semi-conductor defects (see \citealp{LMM2001}). Although such designs are effective for both model selection and estimation for normal-theory linear models, they have been shown to be inefficient for experiments that provide non-normal data \citep{WLER2006}.

In common with other non-linear models, for the GLMs considered in this paper the performance of a design depends on the unknown values of the parameters in the model. One approach to overcoming this problem is to assume a particular value for each parameter and hence obtain a ``locally optimal'' design; that is, a design that is optimal under a given criterion provided the assigned parameter values are correct.  We adopt the alternative approach of making the less stringent assumption of a prior distribution for each model parameter from which we obtain a ``pseudo -Bayesian'' design \citep{AW2015}.

In this paper, we investigate variable screening for GLMs with $q$ independent variables, labelled $x_1, \ldots, x_q$. In the $j$th run $(j = 1,\ldots,N)$ of the experiment, a treatment or combination of variable values $\bx_j = (x_{1j}, \ldots, x_{qj})^\mathrm{T}$ is applied to an experimental unit and a univariate response, $y_j$, is observed. We assume that $|x_{ij}|\le 1$ for $i=1,\ldots,q;\,j=1,\ldots,N$.

The aim of the experiment is to identify those \textit{active} variables having a substantial effect on the response variable and to estimate efficiently a GLM involving those variables alone. For $j=1,\ldots,N$, the $y_j$ have independent exponential family distributions with expectation $\mu_j$ related to a linear predictor $\eta_j = f(\bx_j)^{\mathrm{T}}\bbeta$ via a link function, $g(\mu_j) = \eta_j$. The vectors $f(\bx)$ and $\bbeta$ are $p\times 1$ vectors of known functions of $\bx$ and unknown model parameters, respectively.  We also assume that the experimental units are exchangeable, in the sense that the distribution of the response to a treatment does not depend on the unit to which the treatment is applied.

For canonical link functions, the log-likelihood may be written as
\begin{equation}\label{eq:loglik}
l(\bbeta;\,\by) = \sum_{j=1}^N \left[y_j\eta_j - b(\eta_j) + c(y_j)\right]\,,
\end{equation}
where $b(\cdot)$ and $c(\cdot)$ are known functions of the linear predictor and response, respectively. For the binomial distribution and the logistic link, $b(\eta_j) = -n_j\log(1+e^{\eta_j})$ and $c(y_j) = \log(n_j!/[y_j!(n_j-y_j)!])$, with $n_j$ the number of Bernoulli trials made at the $j$th run. For the Poisson distribution and the log link, $b(\eta_j) = e^{\eta_j}$ and $c(y_j) = -\log(y_j!)$.

Maximum likelihood estimators (MLEs) $\hat{\bbeta}$ can be found via (numerical) maximization of~\eqref{eq:loglik}. For small data sets, however, the MLEs may have considerable bias. For sparse data, such as binomial data with small numbers, $n_j$, of trials for each run, one or more maximum likelihood estimates may be infinite, for example, as the result of separation of the responses into zeros and ones via a hyperplane in the linear predictor \citep{Silvapulle1981}. To remove this bias and guarantee the existence of estimates for GLMs with a canonical link function, \citet{Firth1993} defined penalised maximum likelihood estimators $\tilde{\bbeta}$ as maximisers of
\begin{equation}\label{eq:pl}
l^\star(\bbeta;\,\by) = l(\bbeta;\,\by) + \frac{1}{2}\log \mbox{det}\left\{X^\mathrm{T}WX\right\}\,,
\end{equation}
where $X$ is the $N\times p$ model matrix with $j$th row $f(\bx_j)^{\mathrm{T}}$ and $W = \mathrm{diag}\{\mathrm{var}(y_j)\}$ (see also \citealp{KF2009}). This estimation procedure is equivalent to finding the posterior mode of $\bbeta$ assuming the Jeffreys prior distribution.

The information matrix $X^\mathrm{T}WX$, which is the asymptotic inverse variance-covariance matrix for both $\hat{\bbeta}$ and $\tilde{\bbeta}$, is used to define the $D$-optimality criterion. This criterion specifies selection of a design that maximises the objective function
\begin{equation}\label{eq:dopt}
\phi_D(\xi) = \frac{1}{p}\log \mbox{det}\left\{X^\mathrm{T}WX\right\}\,,
\end{equation}
where
\begin{equation}\label{eq:design} 
\xi = \left\{
\begin{array}{ccc}
\bx_1 & \ldots & \bx_n \\
\omega_1 & \ldots & \omega_n \\
\end{array}
\right\}\,,
\end{equation}
$\bx_1,\ldots,\bx_n$ are the distinct treatments in the design (assumed, without loss of generality, to be applied to the first $n$ runs of the experiment), $\omega_k>0\in\mathbb{N}$, and $\sum_{k=1}^n\omega_k = N$, the total number of runs. For the GLMs considered in this paper,~\eqref{eq:dopt} depends on $\bbeta$ through the matrix $W$ and hence selection of a $D$-optimal design requires knowledge of the values of these parameters. Thus a locally optimal design is obtained.

The relative performance of two designs, $\xi_1$ and $\xi_2$, under $D$-optimality may be assessed using relative $D$-efficiency, defined as
\begin{equation}\label{eq:Deff}
\mbox{DEff}(\xi_1,\xi_2) = \exp\left\{\phi_D(\xi_1) - \phi_D(\xi_2)\right\}\,,
\end{equation}
where $0\le\mbox{DEff}(\xi_1,\xi_2)$. If $\xi_2$ is a $D$-optimal design that maximises~\eqref{eq:dopt}, then~\eqref{eq:Deff} provides an absolute measure of the performance of design $\xi_1$.

In this paper, we address the screening problem of model selection and estimation of parameters in the selected model. We define, in Section~\ref{sec:methods}, a Bayesian information capacity criterion that generalises $D$-optimality to provide model-robust designs for GLMs. We also present and discuss a model selection strategy that uses all-subsets regression and suitable penalties for model complexity. Sections~\ref{sec:binomial} and~\ref{sec:poisson} describe simulation studies of logistic and log-linear regression modelling, respectively, which demonstrate and assess the effectiveness of the methods. In Section~\ref{sec:disc}, we present some avenues for future work to further develop methodology for screening experiments with non-normal data.

\section{Information capacity designs and model selection}\label{sec:methods}

Consider a set $\mathcal{M}$ of $M = |\mathcal{M}|$ distinct candidate models, each of which have the same link function. The linear predictor for the $m$th model and $j$th run is given by 
\begin{equation}\label{eq:IClinpred}
\eta_j^{m} = \beta_{0m} + \sum_{i = 1}^q \beta_{im}x_{ij}I(i, m)\,,
\end{equation}
where the $\beta_{im}$ are the values of the parameters in model $m$, $I(i,m)=1$ if variable $i$ is in model $m$, and $I(i,m)=\beta_{im}=0$ otherwise. Hence, the number of parameters in model $m$ is $p_m = 1 + \sum_{i=1}^q I(i,m)$.

\subsection{Bayesian information capacity}\label{sec:ic}

Information capacity (IC) was introduced as a linear-model design selection criterion by \citet{sun1993}. It has been further developed and applied by, for example, \citet{wu1993} (supersaturated designs), and \citet{LN2000} (model-robust factorial designs). In essence, this criterion seeks a design whose projections onto subsets of the variables produce sub-designs having good estimation properties for the corresponding submodels. This is achieved by selecting a design that maximises a weighted average of the $D$-criterion objective function for each submodel.

For GLMs, \citet{woods2010} employed the criterion of \citet{WLER2006} to find locally optimal information capacity designs for an example having five variables. Designs were found that maximised
\begin{equation}\label{eq:locoptIC}
\Psi(\xi) = \sum_{m = 1}^M \frac{1}{p_m}\log\mbox{det}\left\{ X_m^\mathrm{T} W_m X_m\right\}\,,
\end{equation}
where $X_m$ and $W_m$ are the respective model and weight matrices for the $m$th model in $\mathcal{M}$.

We define the Bayesian IC criterion which incorporates into~\eqref{eq:locoptIC} uncertainty in the parameter values assumed for each model. This criterion selects a design that maximises the objective function
\begin{equation}\label{eq:BayesIC}
\Phi(\xi) = \sum_{m = 1}^M \frac{1}{p_m}\int _{\mathcal{B}_m} \log\mbox{det}\left\{ X_m^\mathrm{T} W_m X_m\right\}\pi_m(\bbeta_m)\,\mathrm{d}\bbeta_m\,,
\end{equation}
where $\mathcal{B}_m\subset\mathbb{R}$ is the parameter space for model $m$, $\bbeta_m = (\beta_{0m}, \ldots, \beta_{qm})^\mathrm{T}$, and $\pi_m(\bbeta_m)$ is the prior distribution for $\bbeta_m$. The choice of $\pi(\bbeta_m)$ and $\mathcal{B}_m$, and the evaluation of~\eqref{eq:BayesIC}, are discussed in Section~\ref{sec:binic} for logistic regression. For log-linear regression, we make use of results in the literature that enable analytical construction of minimally-supported $D$- and Bayesian $D$-optimal designs, see Section~\ref{sec:poisic}.

\subsection{Model selection}\label{sec:ms}

A variety of model selection procedures exist for determining the most appropriate GLM from a set of models, including Bayesian \citep{CHIK2008} and shrinkage methods \citep{PH2007}. To focus investigations on the impact of design selection, we restrict attention to all-subsets regression and use an information criterion to adjust for the bias inherent from in-sample estimation of the prediction error (see \citealp{BA2002}, ch. 2). When maximum likelihood estimation is employed, we use the Akaike information criterion (AIC; \citealp{aic_1974}) as the model selection criterion, and choose a model that minimises
\begin{equation}\label{eq:aic}
AIC(m;\, \hat{\bbeta}) = -2l_m(\hat{\bbeta};\,\by) + 2p_m\,,
\end{equation}
where $l_m(\cdot;\cdot)$ is the log-likelihood function~\eqref{eq:loglik} for model $m$. 

When $\bbeta$ is estimated via penalised maximum likelihood (see~\eqref{eq:pl}), AIC is no longer an appropriate criterion. This is because the effective number of parameters is reduced, equivalent to the inclusion of prior information (\citealp{GHV2014}). The reduction depends on the number ($N$) of runs, with a smaller number of effective parameters for smaller $N$. Hence when $N$ is small, use of AIC will over-penalise larger models. To avoid this problem, we use a generalised information criterion (GIC; \citealp{KK1996}) that relaxes the assumptions of (i) estimation via maximum likelihood, and (ii) inclusion of the true model in $\mathcal{M}$. Hence, we select the model that minimises
\begin{equation}\label{eq:gic}
GIC(m;\,\tilde{\bbeta}) = -2l_m(\tilde{\bbeta};\, \by) + 2\tr\left\{J^{-1}(\tilde{\bbeta})I(\tilde{\bbeta})\right\}\,,
\end{equation}
where
\begin{equation*}
J(\tilde{\bbeta}) = -\frac{1}{N}\,\left.\frac{\partial^2 l_m^\star(\bbeta;\,\by)}{\partial\bbeta\partial\bbeta^\mathrm{T}}\right|_{\tilde{\bbeta}}\,,
\quad I(\tilde{\bbeta}) = \frac{1}{N}\, \left.\frac{\partial l_m^\star(\bbeta;\,\by)}{\partial\bbeta}\frac{\partial l_m(\bbeta;\,\by)}{\partial\bbeta^\mathrm{T}}\right|_{\tilde{\bbeta}}\,,
\end{equation*}
with $l_m^\star(\cdot;\cdot)$ the penalised log-likelihood function~\eqref{eq:pl} for model $m$; see also \citet{MYA1994} and \citet{ZLT2010}. The evaluation of $J(\tilde{\bbeta})$ and $I(\tilde{\bbeta})$ is straightforward for the GLMs and penalised likelihood estimation method used in this paper. The performance of the GIC is investigated in Section~\ref{sec:binresults}.

Following analysis of the data from an experiment, those variables found to be involved in the selected model are deemed to be active. In simulation studies to assess the performance of the model selection strategies, we use three summary measures: (i) \textit{power}: the proportion of truly active variables that are correctly identified as active by the model selection strategy; (ii) \textit{type I error rate}: the proportion of inactive variables (i.e. those not included in the true model) that are incorrectly identified as active by the model selection strategy; and (iii) \textit{false discovery rate} (FDR): the proportion of variables identified as active by the model selection strategy that are truly inactive (i.e. not in the true model). 

\section{Designs and model selection for binomial response and logistic regression}\label{sec:binomial}

To investigate the performance of the methodology for logistic regression we study a five variable example, with linear predictors of the form~\eqref{eq:IClinpred}. We assume that any subset of these variables may be the set of active variables. Therefore there are 31 possible models. The models are ordered lexicographically within each model size and assigned labels $1,\ldots,31$. Models $1,\ldots,5$ have linear predictors that contain a single variable, $1,\ldots,5$, respectively; models $6,\ldots,15$ have two variables, $1,2;\,1,3;\,\ldots,4,5$. Similarly, models $16,\ldots,25$ are three variable models, 26 - 30 are four-variable models and model 31 contains all five variables. 

To find optimal designs and perform subsequent simulation studies, the model parameters $\beta_{im}$ are assumed to have independent prior distributions of the form
\begin{equation}\label{eq:prior}
\beta_{im} \sim \left\{
\begin{array}{ll}
\mbox{Uniform}(\kappa, 5) & \mbox{for } i = 1,3,4 \mbox{ and } I(i, m) = 1\,, \\
\mbox{Uniform}(-5, -\kappa) & \mbox{for } i = 2,5 \mbox{ and } I(i, m) = 1\,, \\
\end{array}
\right.
\end{equation}
where $\kappa = 1,2,3$ and we assume $\beta_{0m} = 0$ and $\beta_{im}=0$ if $I(i,m)=0$ for $m = 1, \ldots, M$. The adoption of bounded uniform prior distributions prevents the occurrence of parameter vectors in the support of the prior for which no design has a non-singular information matrix (c.f. \citealp{waite2015}).

\subsection{Information capacity designs}\label{sec:binic}

\newcommand{\bindeffscale}{.3}
\begin{figure}
\begin{center}
\begin{tabular}{>{\centering\arraybackslash}m{.9cm}>{\centering\arraybackslash}m{5cm}>{\centering\arraybackslash}m{6cm}}
 & Bayesian IC & Locally $D$-optimal \\[-2ex] 
$\kappa = 1$ & \includegraphics[scale=\bindeffscale]{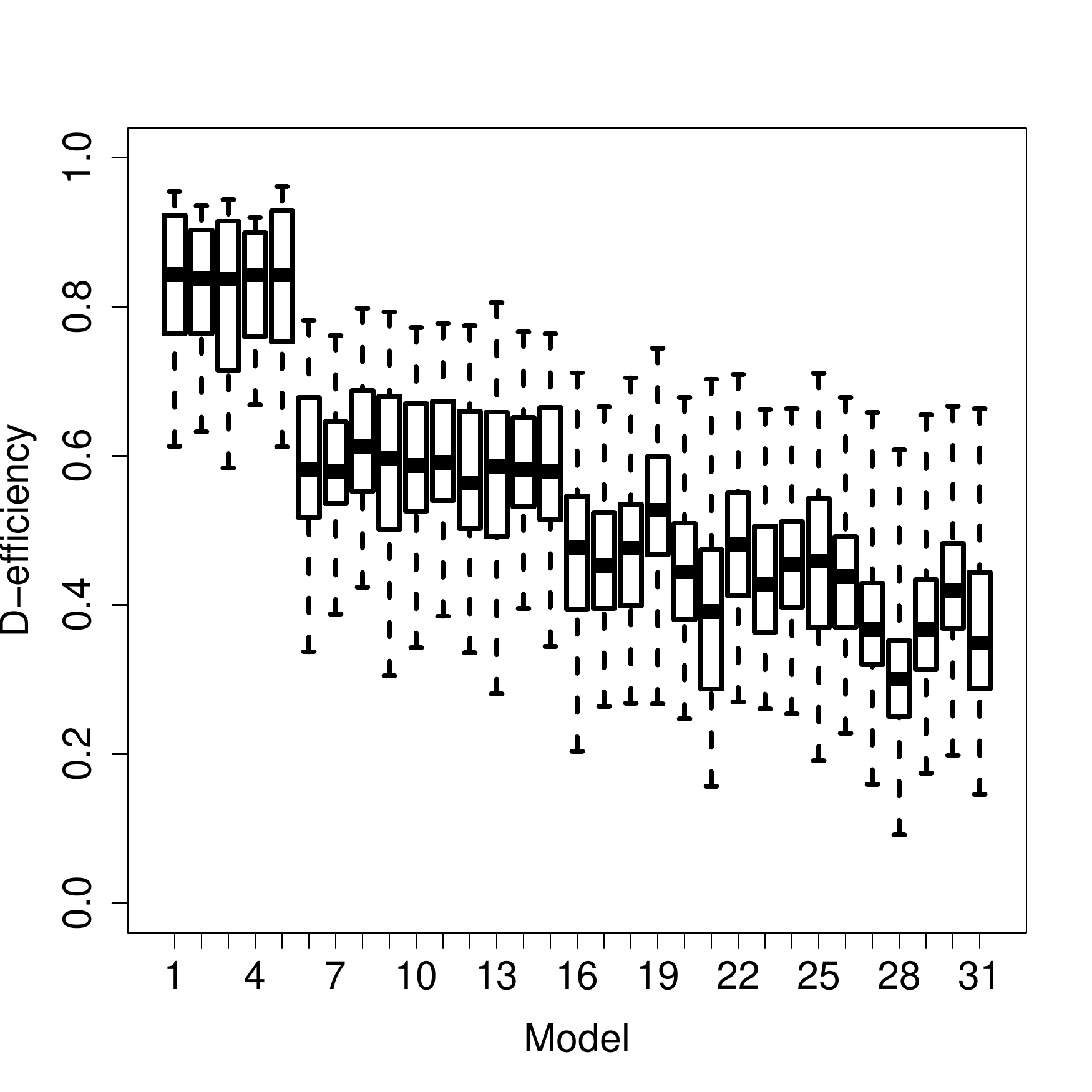} & \includegraphics[scale=\bindeffscale]{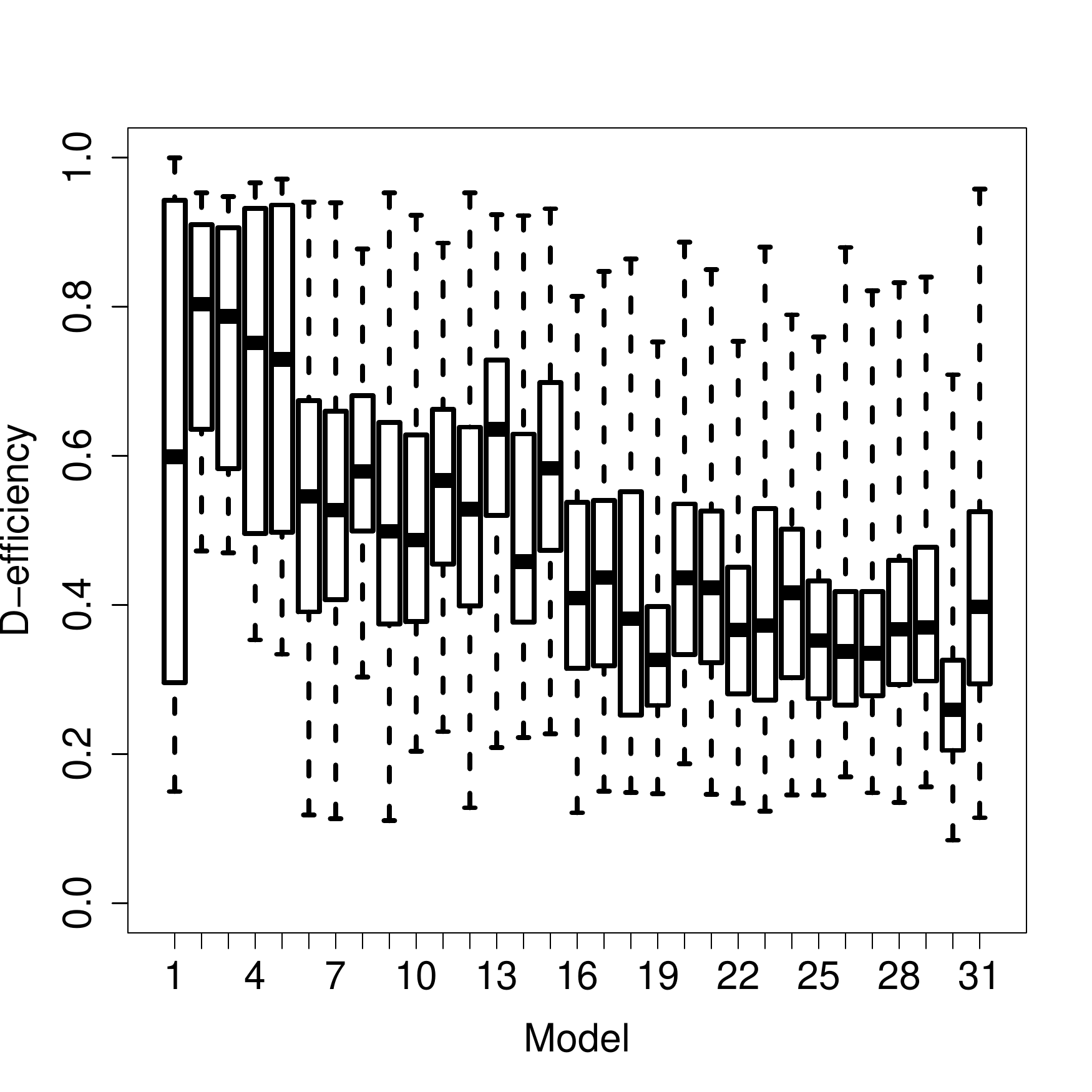} \\  
$\kappa = 2$ & \includegraphics[scale=\bindeffscale]{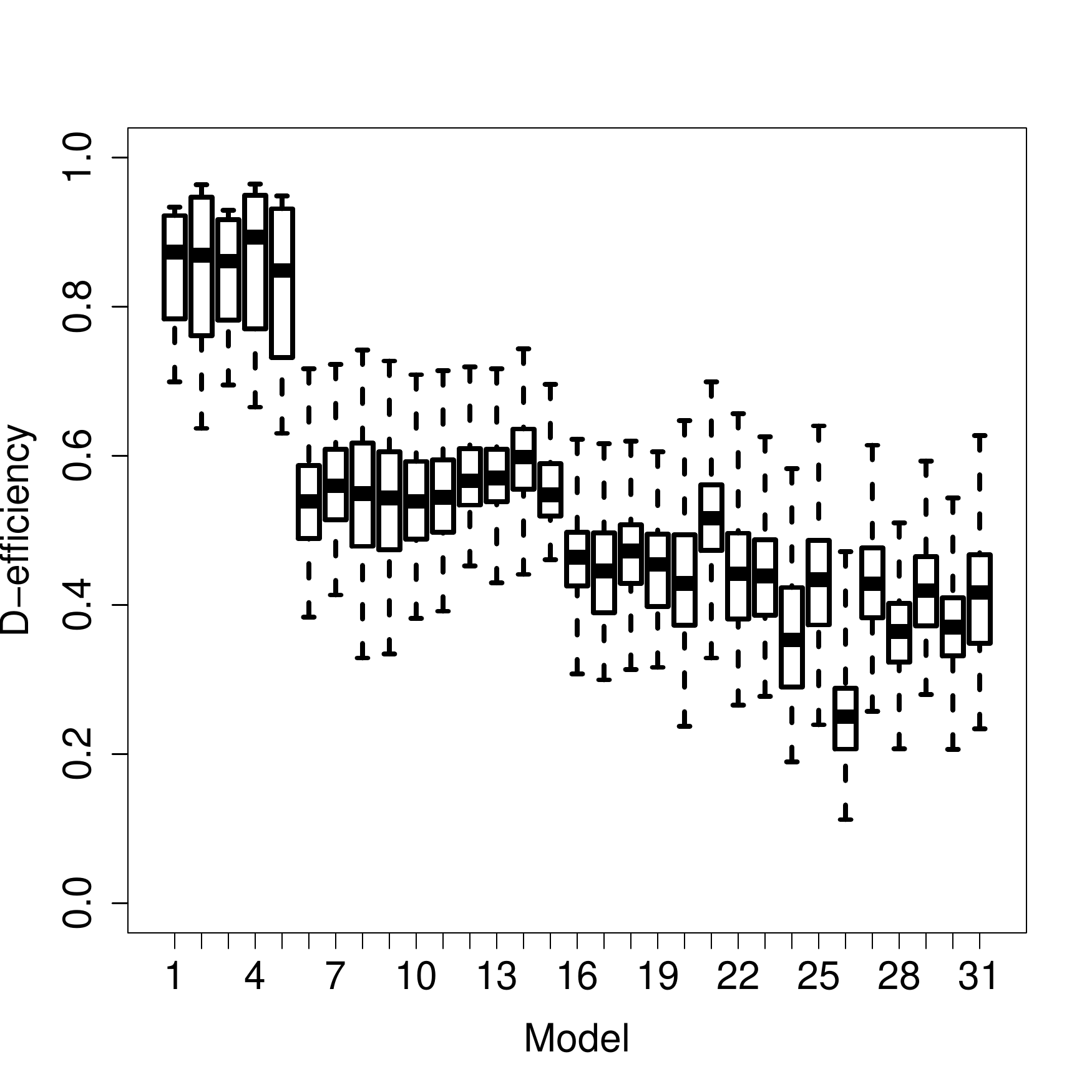} & \includegraphics[scale=\bindeffscale]{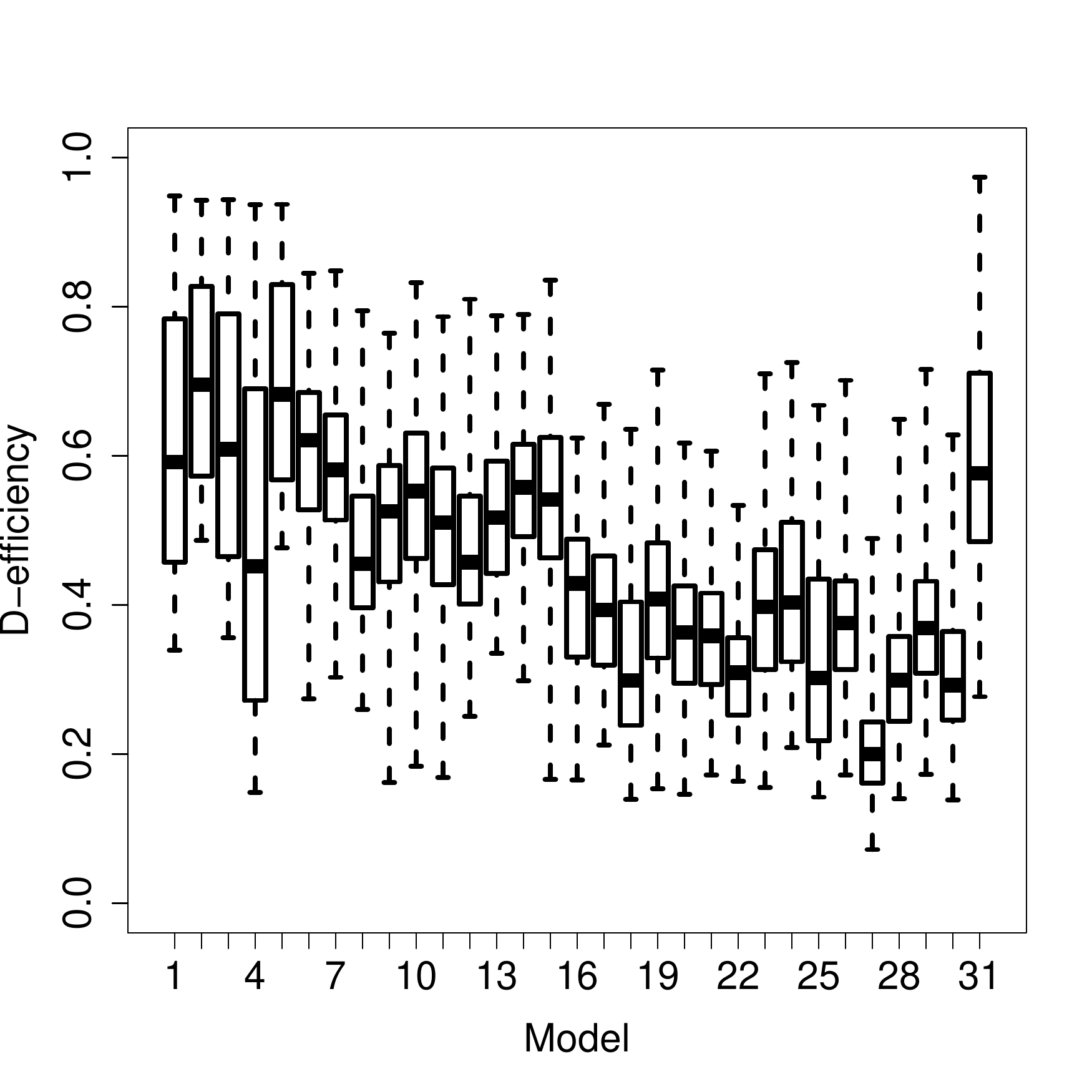} \\
$\kappa = 3$ & \includegraphics[scale=\bindeffscale]{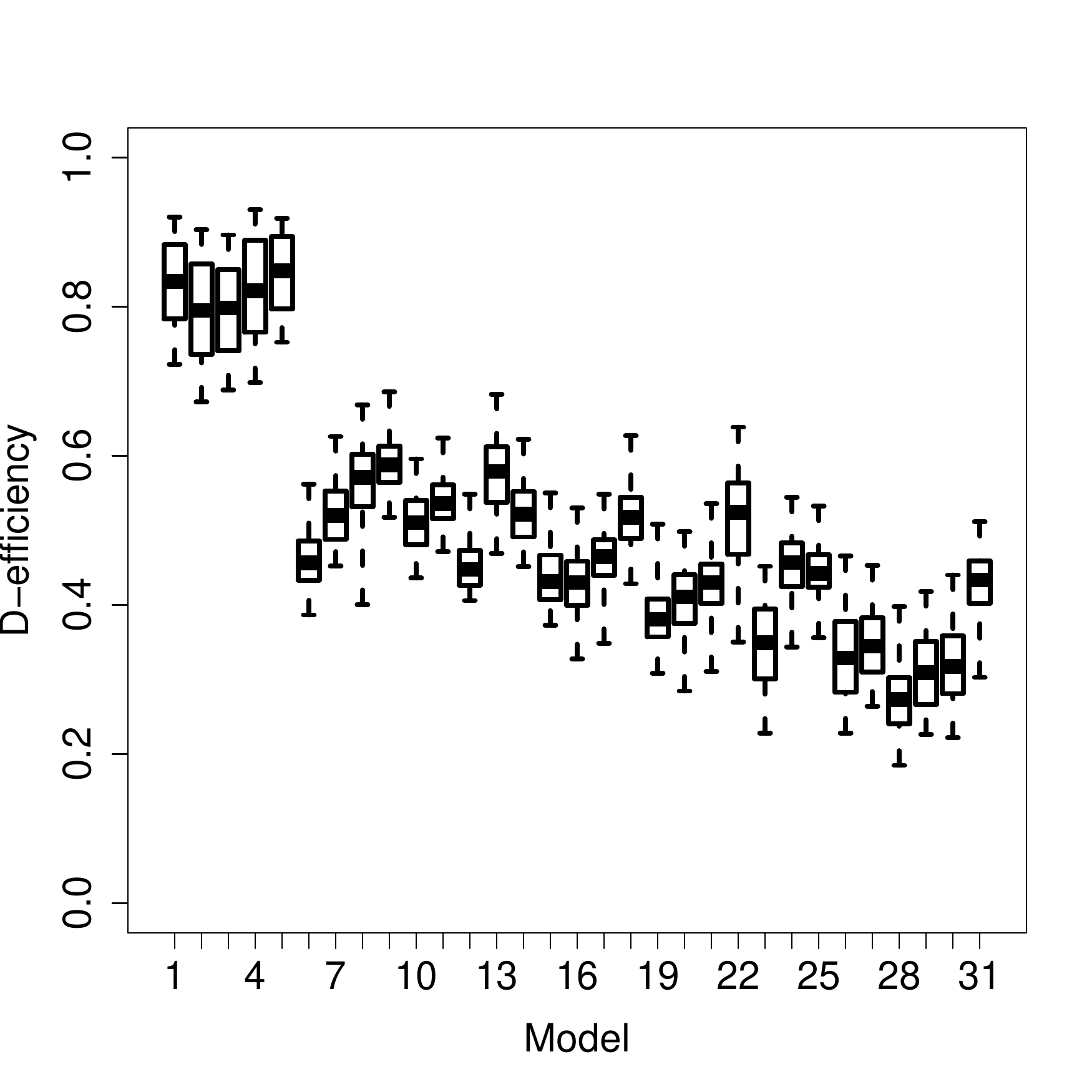} & \includegraphics[scale=\bindeffscale]{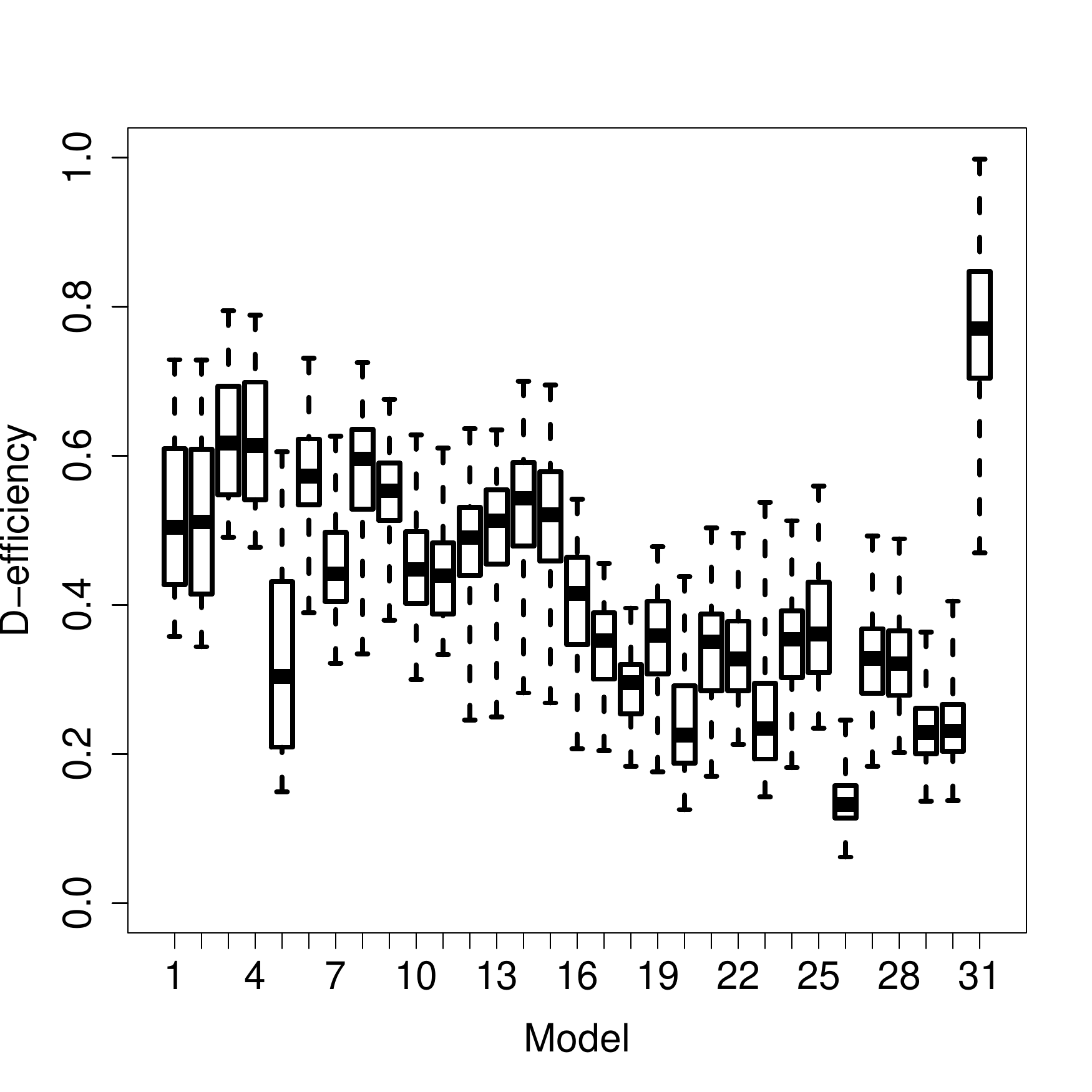} \\
\end{tabular}
\end{center}
\caption{\label{fig:deff6}Boxplots of $D$-efficiencies for Bayesian information capacity designs (left column) and locally $D$-optimal designs for the maximal model (right column) for logistic regression with $n=6$ support points and three prior distributions ($\kappa=1,2,3$) for the model parameters.}
\end{figure}

\begin{figure}
\begin{center}
\begin{tabular}{>{\centering\arraybackslash}m{.9cm}>{\centering\arraybackslash}m{5cm}>{\centering\arraybackslash}m{6cm}}
 & Bayesian IC & Locally $D$-optimal \\[-2ex] 
$\kappa = 1$ & \includegraphics[scale=\bindeffscale]{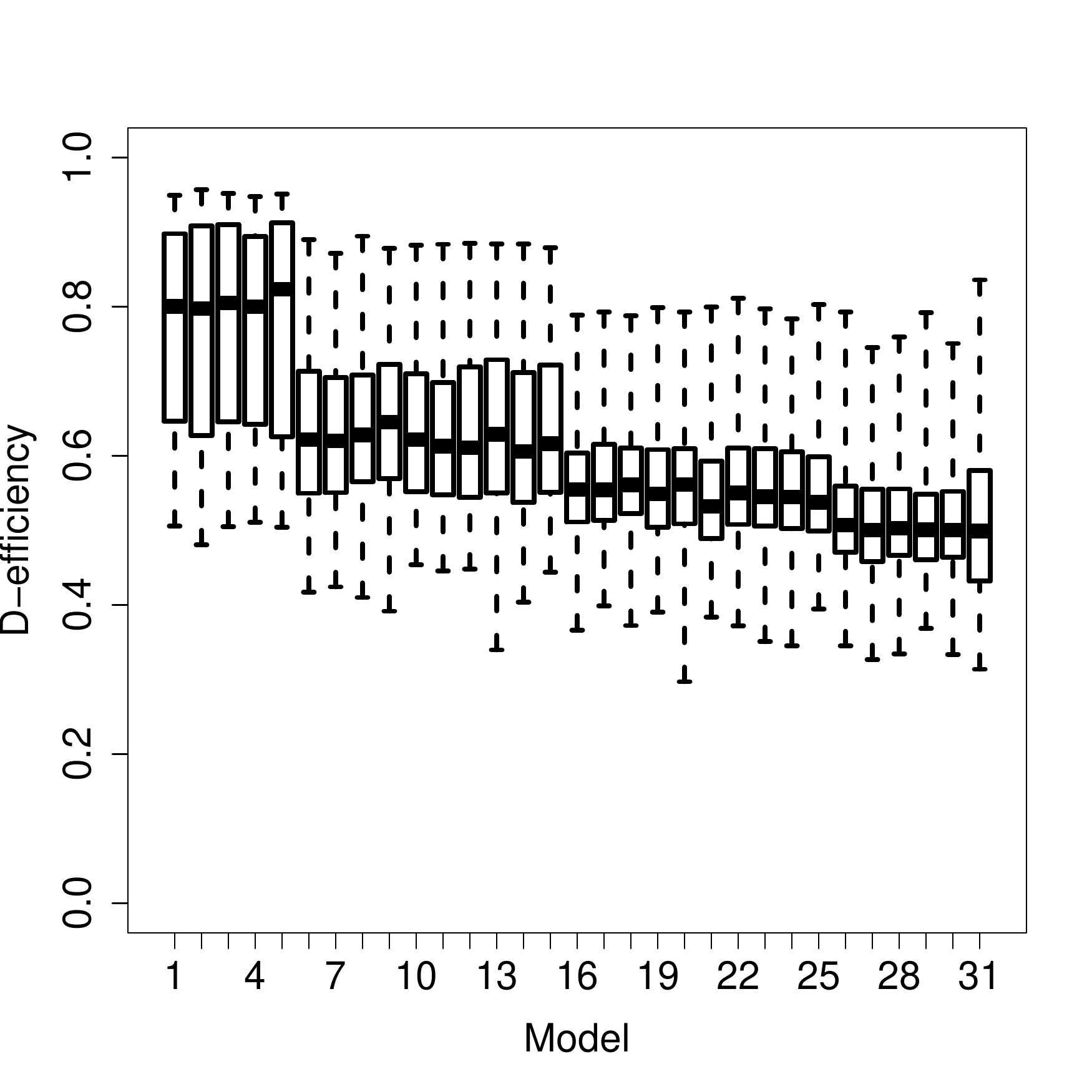} & \includegraphics[scale=\bindeffscale]{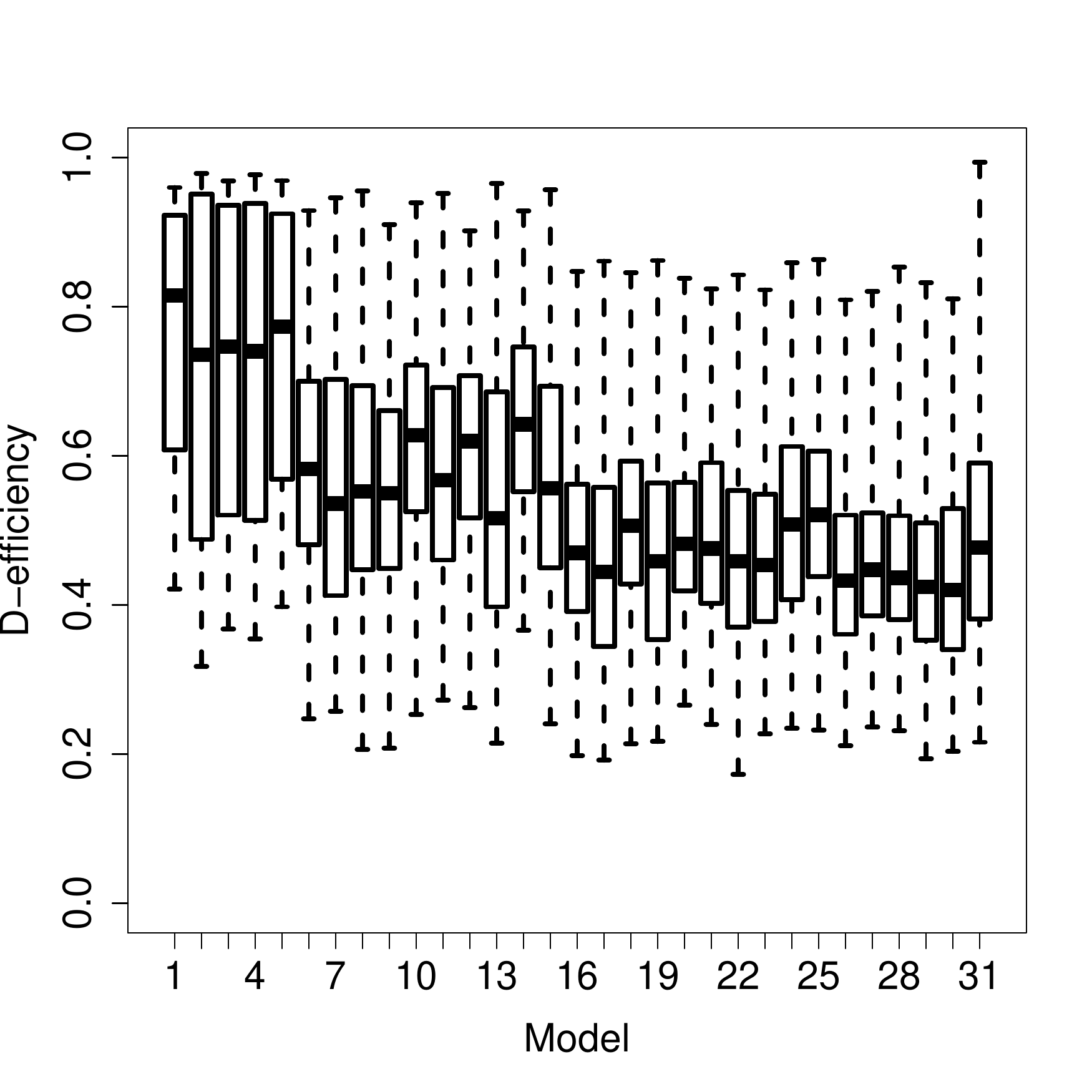} \\  
$\kappa = 2$ & \includegraphics[scale=\bindeffscale]{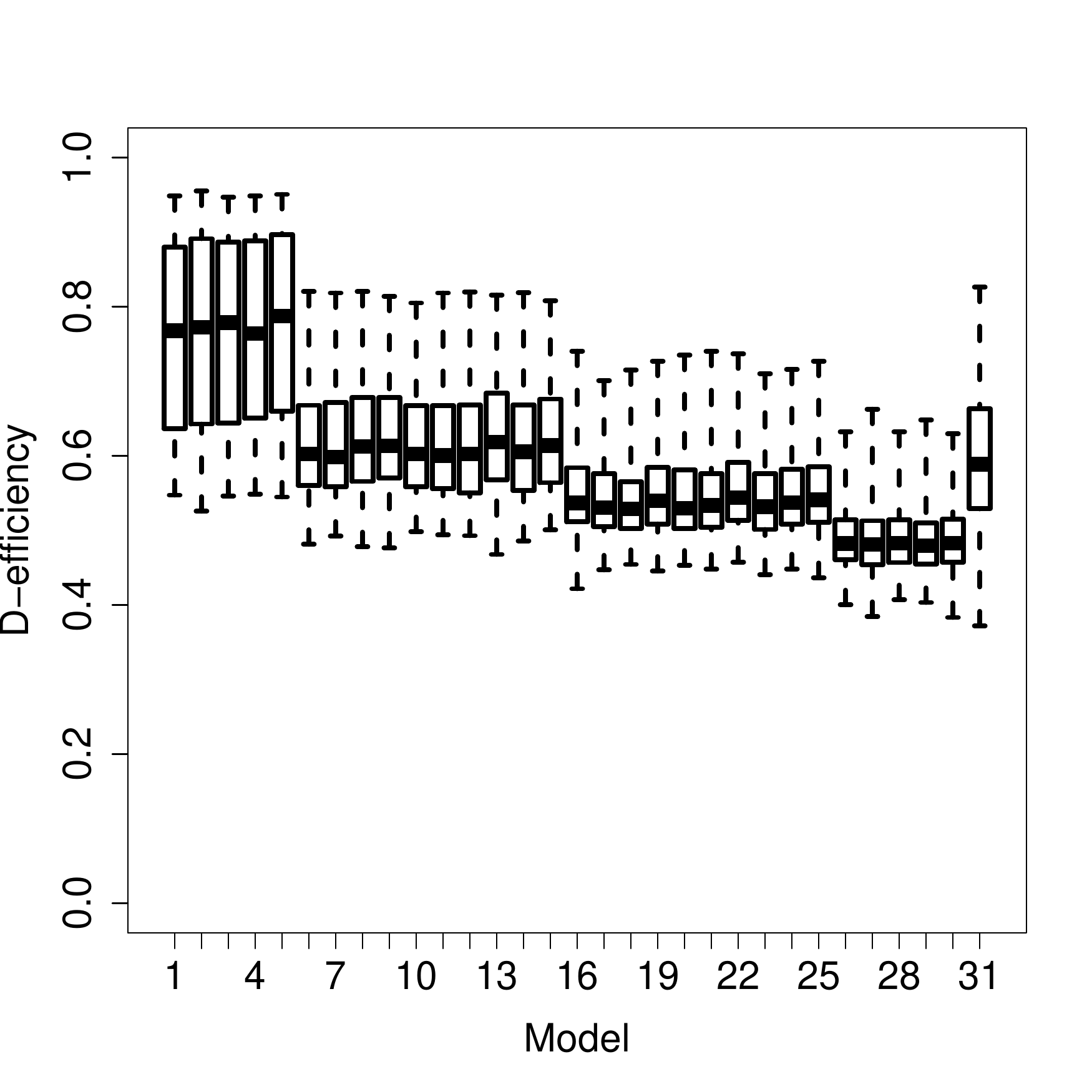} & \includegraphics[scale=\bindeffscale]{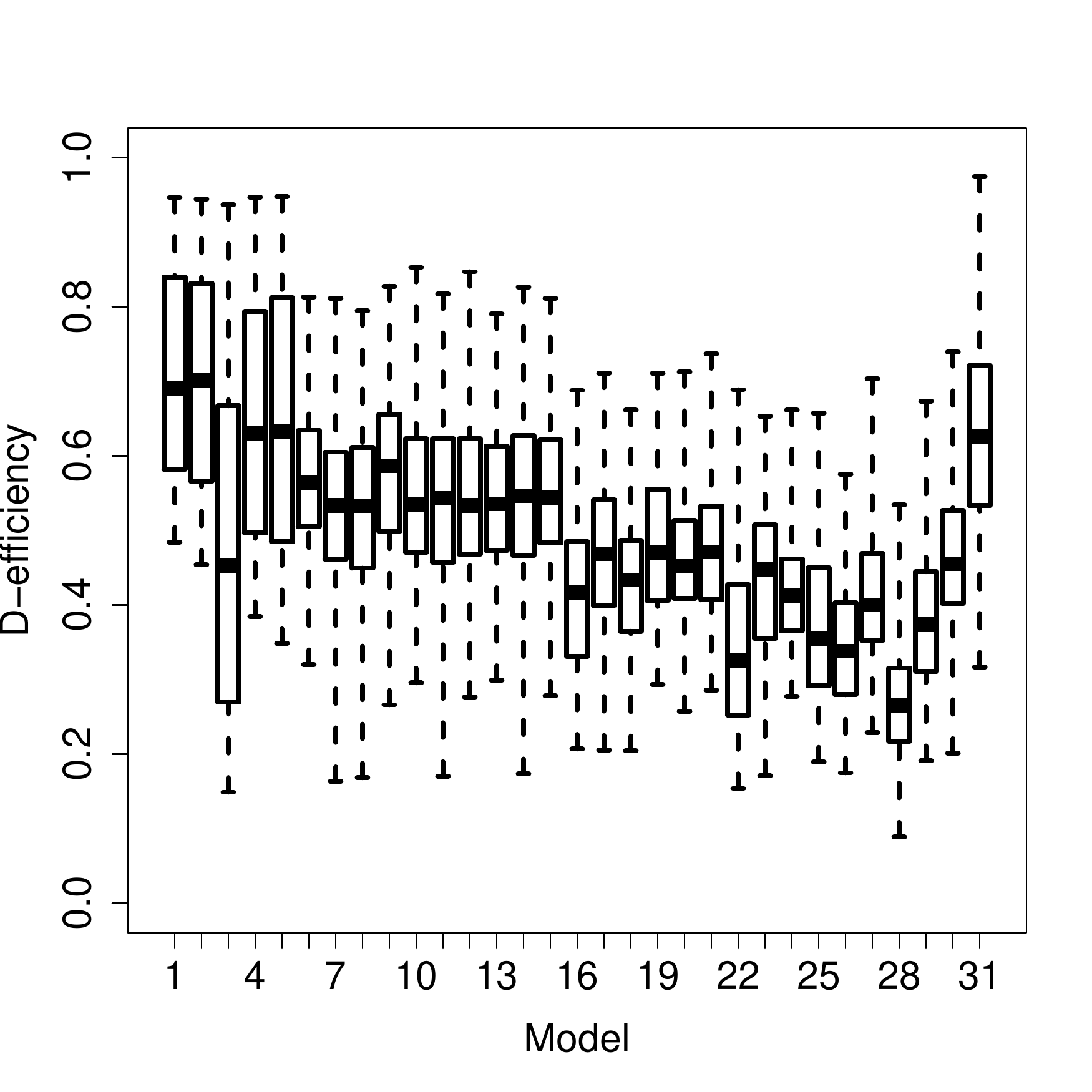} \\
$\kappa = 3$ & \includegraphics[scale=\bindeffscale]{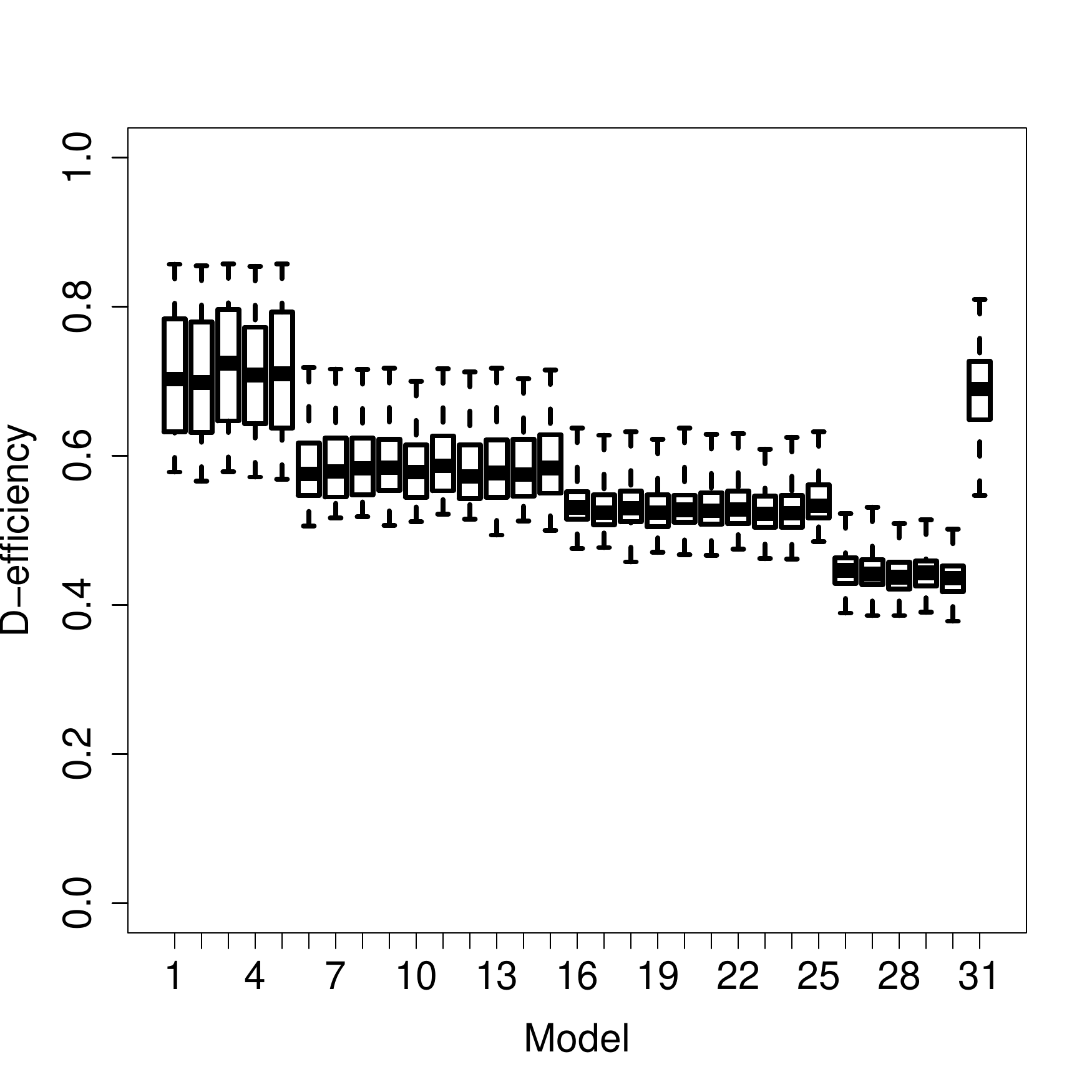} & \includegraphics[scale=\bindeffscale]{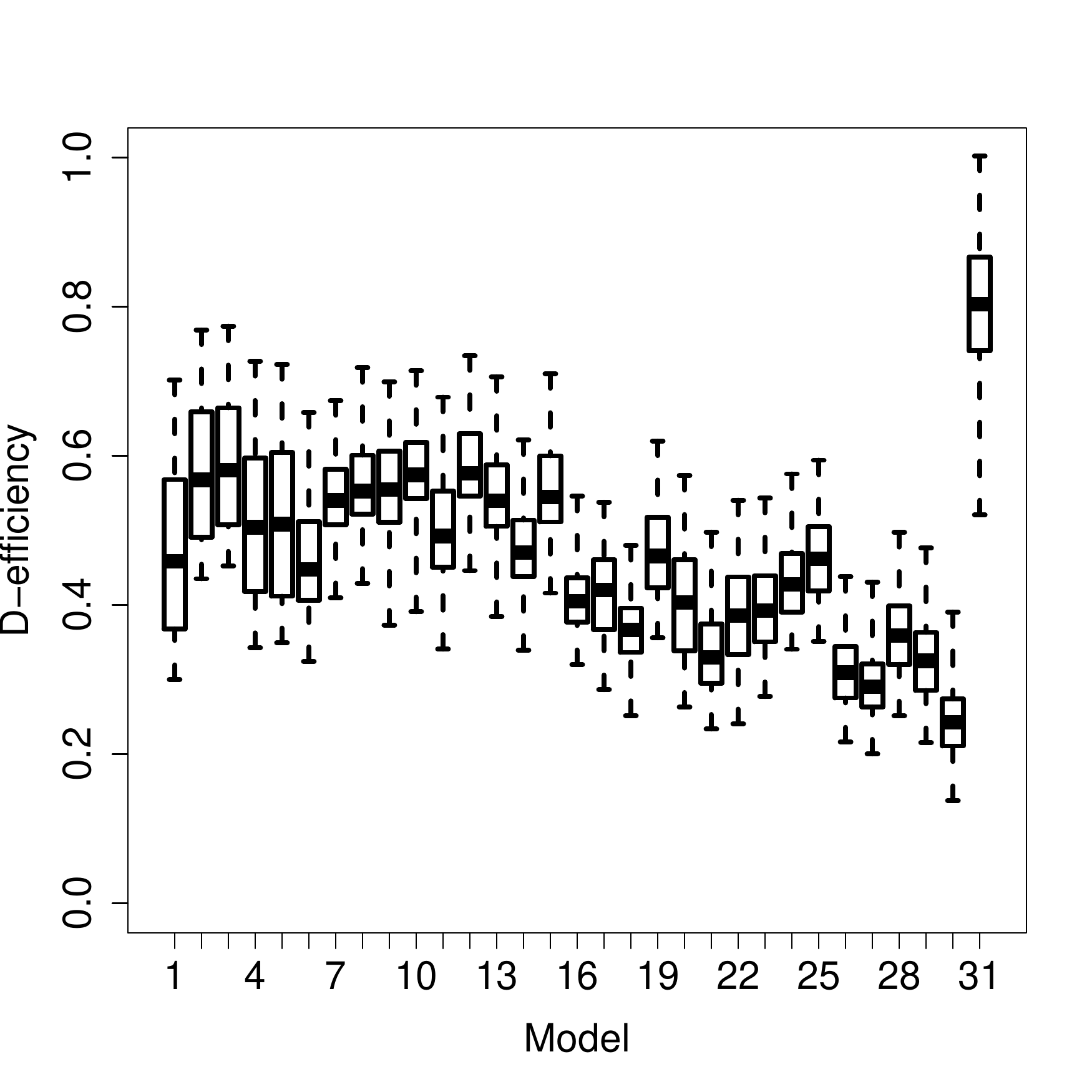} \\
\end{tabular}
\end{center}
\caption{\label{fig:deff30}Boxplots of $D$-efficiencies for Bayesian information capacity designs (left column) and locally $D$-optimal designs for the maximal model (right column) for logistic regression with $n=30$ support points and three prior distributions ($\kappa=1,2,3$) for the model parameters.}
\end{figure}

We relax the assumption in~\eqref{eq:design} that $\omega_k$ is integer ($k = 1,\ldots,n$) and consider approximate designs (e.g. \citealp{ADT}, ch. 9). An approximate Bayesian IC design for logistic regression maximises
\begin{equation}\label{eq:icapprox}
\Phi^\dagger(\xi) = \sum_{m = 1}^M \frac{1}{p_m}\int _{\mathcal{B}_m} \log\mbox{det}\left\{ \sum_{k=1}^n \tilde{\omega}_k\mathrm{var}(y_k)f_m(\bx_k)f_m(\bx_k)^\mathrm{T}\right\}\pi_m(\bbeta_m)\,\mathrm{d}\bbeta_m\,,
\end{equation}
where $0<\tilde{\omega}_k = \omega_k/N\le 1$ and $f_m(\bx_k)^{\mathrm{T}}\bbeta_m$ is the linear predictor for the $m$th model. Clearly, an optimal choice of approximate Bayesian IC design can be made independently of the total experiment size $N$. Finding approximate designs also substantially reduces the computational burden of the design optimisation. We found designs using simulated annealing \citep{haines1987} where the integral in~\eqref{eq:icapprox} was evaluated numerically as a summation across a quasi-Monte Carlo sample \citep[][ch. 5]{Lemieux2009}. The simulated annealing algorithm employed was a cyclic descent algorithm that proposed, evaluated and accepted moves for one coordinate of the design at a time; see \citet{woods2010}. Such ``coordinate exchange'' algorithms are standard in the design of experiments, and solve difficult, high-dimensional, optimisation problems via a series of one-dimensional optimisations. Use of a stochastic optimisation algorithm such as annealing has the advantage of helping to escape local optima which we have found is a particular issue when each coordinate can take values in a continuous range.

Figures~\ref{fig:deff6} and~\ref{fig:deff30} (left columns) summarise the $D$-efficiencies of the Bayesian IC designs for $n=6$ and $n=30$ support points, respectively. For each choice of prior distribution for $\kappa=1,2,3$ from~\eqref{eq:prior}, the plots are obtained using (i) 500 random draws of $\bbeta_m$ values from distribution~\eqref{eq:prior}; (ii) the locally $D$-optimal design for each value of $\bbeta_m$, again found using simulated annealing; and (iii) calculation of efficiency~\eqref{eq:Deff} to compare the Bayesian IC design with the locally $D$-optimal design. In general, the efficiencies decrease with model size. For $n=6$, the efficiencies are highly variable between models, even for models of the same size. It is not uncommon for a Bayesian design for logistic regression to require a large number of support points (see \citealp{cl1989}, and \citealp{wl2011}). This variability is also evident in the simulation results for model selection. Hence, in the next section, we present only assessments of the designs with $n=30$ support points.

For comparison, Figures~\ref{fig:deff6} and~\ref{fig:deff30} (right columns) also presents the $D$-efficiencies of locally $D$-optimal designs for the maximal model containing all five variables (model 31) and with each parameter set to its prior expectation. With the obvious, and expected, exception of model 31, the $D$-efficiencies are more variable and generally lower than those obtained from the Bayesian IC design. 

\subsection{Model selection results}\label{sec:binresults}

To assess the performance of the designs for model selection, a simulation study was performed for each Bayesian IC design in which (i) each of the models, in turn, was used as the true model for the data generating process; (ii) 1000 data sets were generated independently by simulating values of $\bbeta_m$ from the prior distribution, followed by simulation of responses $\by$ from a Binomial distribution; (iii) for each data set, each of the models in $\mathcal{M}$ was fitted using maximum penalised likelihood, and the model selected that minimises GIC~\eqref{eq:gic}; and (iv) power, type I error rate and FDR were calculated for each simulated data set. The optimal approximate designs were converted into exact designs via rounding $\omega_k$ to the nearest integer. Results are provided for $N = 30, 50, 80, 100$ in Figures~\ref{fig:binsim1} and~\ref{fig:binsim3}, for $\kappa = 1,3$ respectively. The results for $\kappa = 2$ (not shown) are similar to those for $\kappa=3$.   

\newcommand{\binsimscale}{.33}
\begin{figure}
\begin{center}
\begin{tabular}{cc}
$N = 30$ & $N = 50$ \\[-2ex]
\includegraphics[scale=\binsimscale]{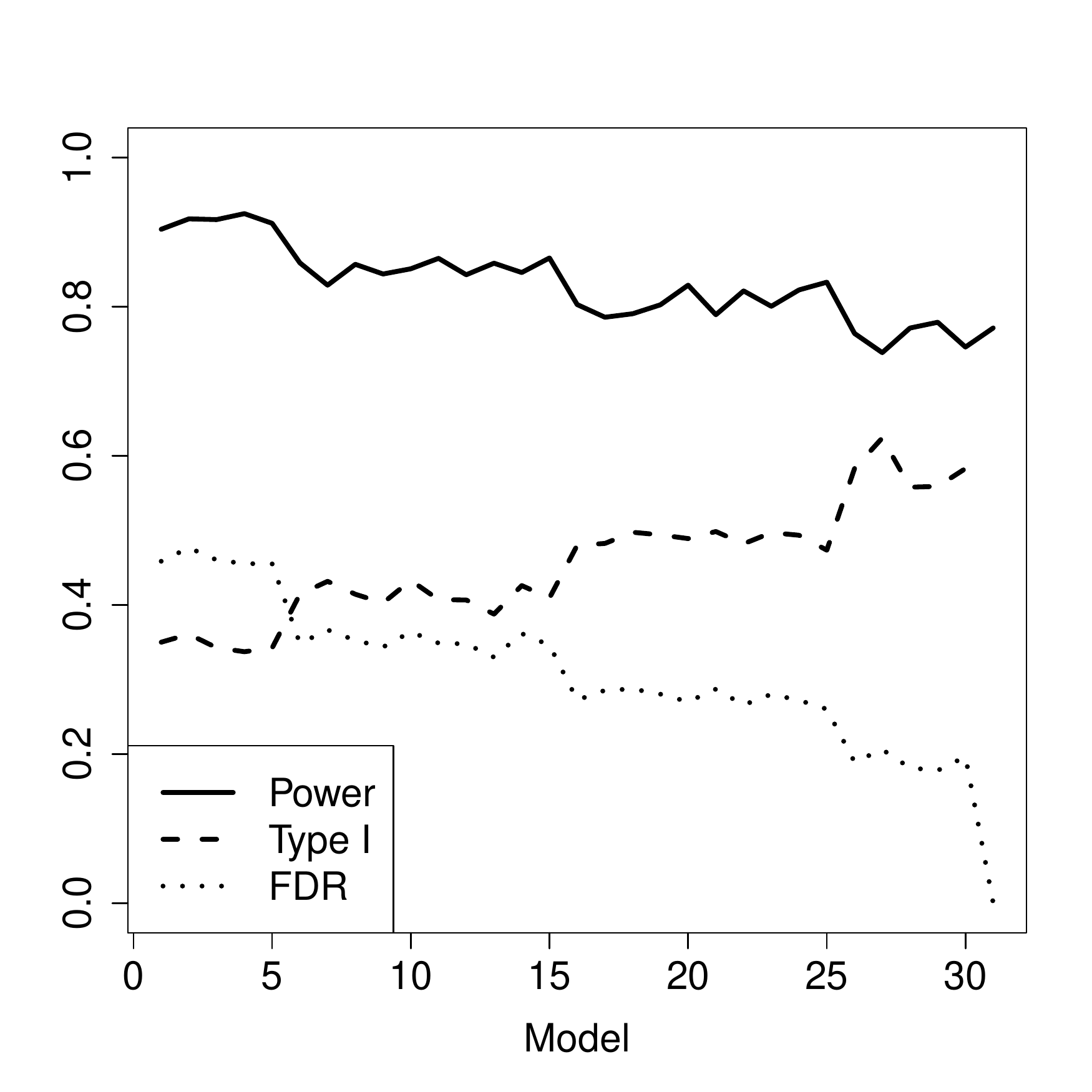} & \includegraphics[scale=\binsimscale]{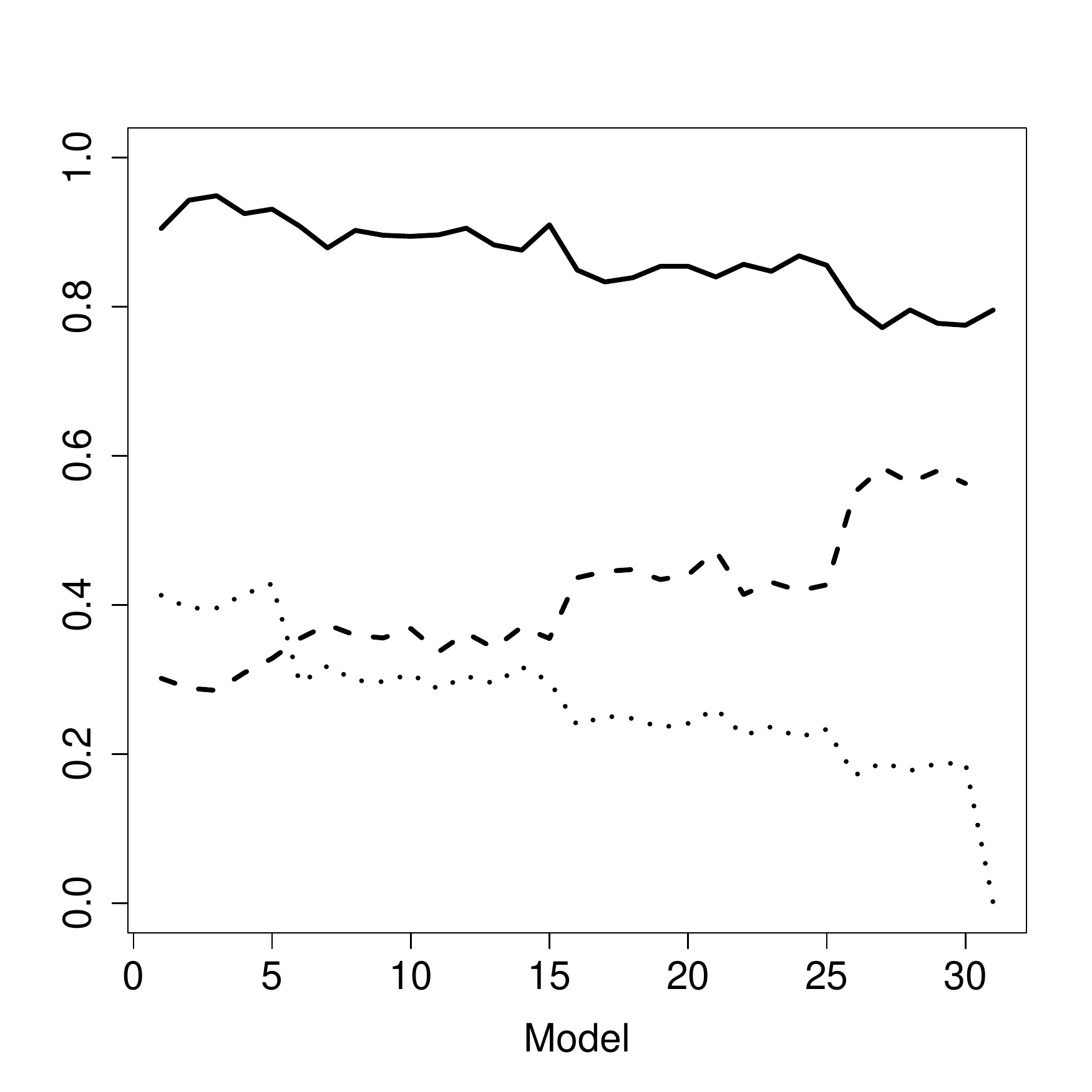} \\
$N = 80$ & $N = 100$ \\[-2ex]
\includegraphics[scale=\binsimscale]{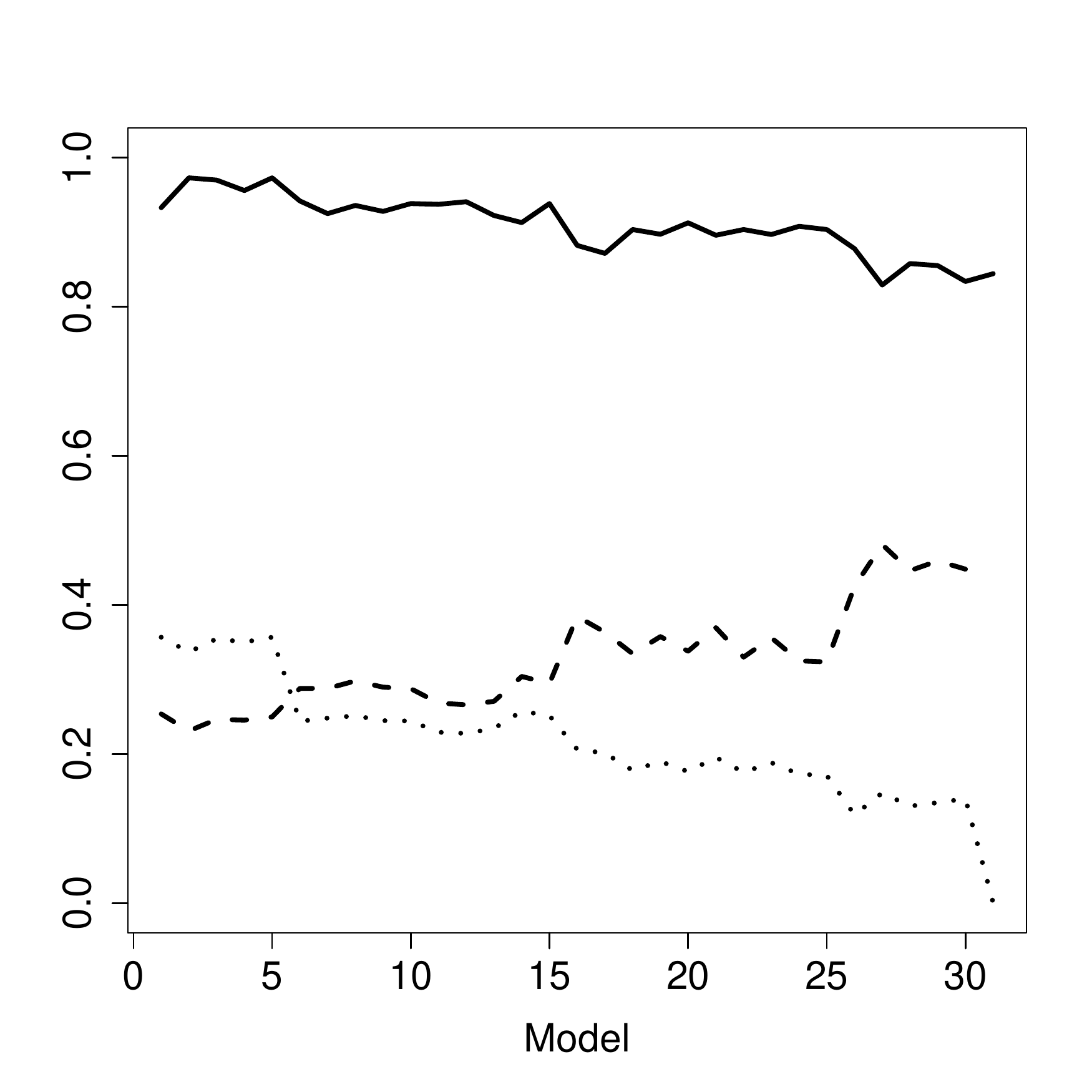} & \includegraphics[scale=\binsimscale]{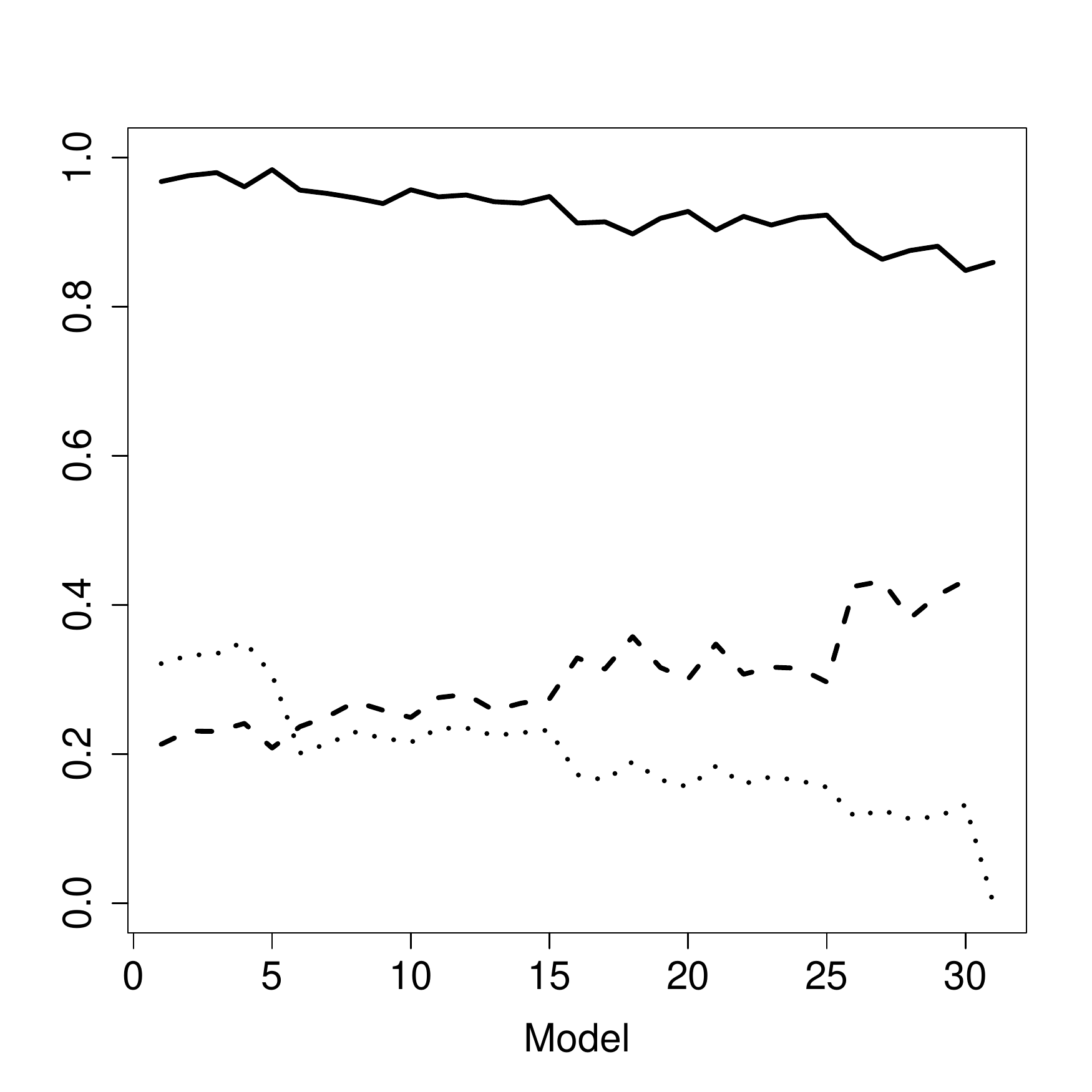}
\end{tabular}
\end{center}
\caption{\label{fig:binsim1}Average power, type I error rate and FDR for logistic regression with $\kappa = 1$ and $N=30,50,80,100$ runs.}
\end{figure}


\begin{figure}
\begin{center}
\begin{tabular}{cc}
$N = 30$ & $N = 50$ \\[-2ex]
\includegraphics[scale=\binsimscale]{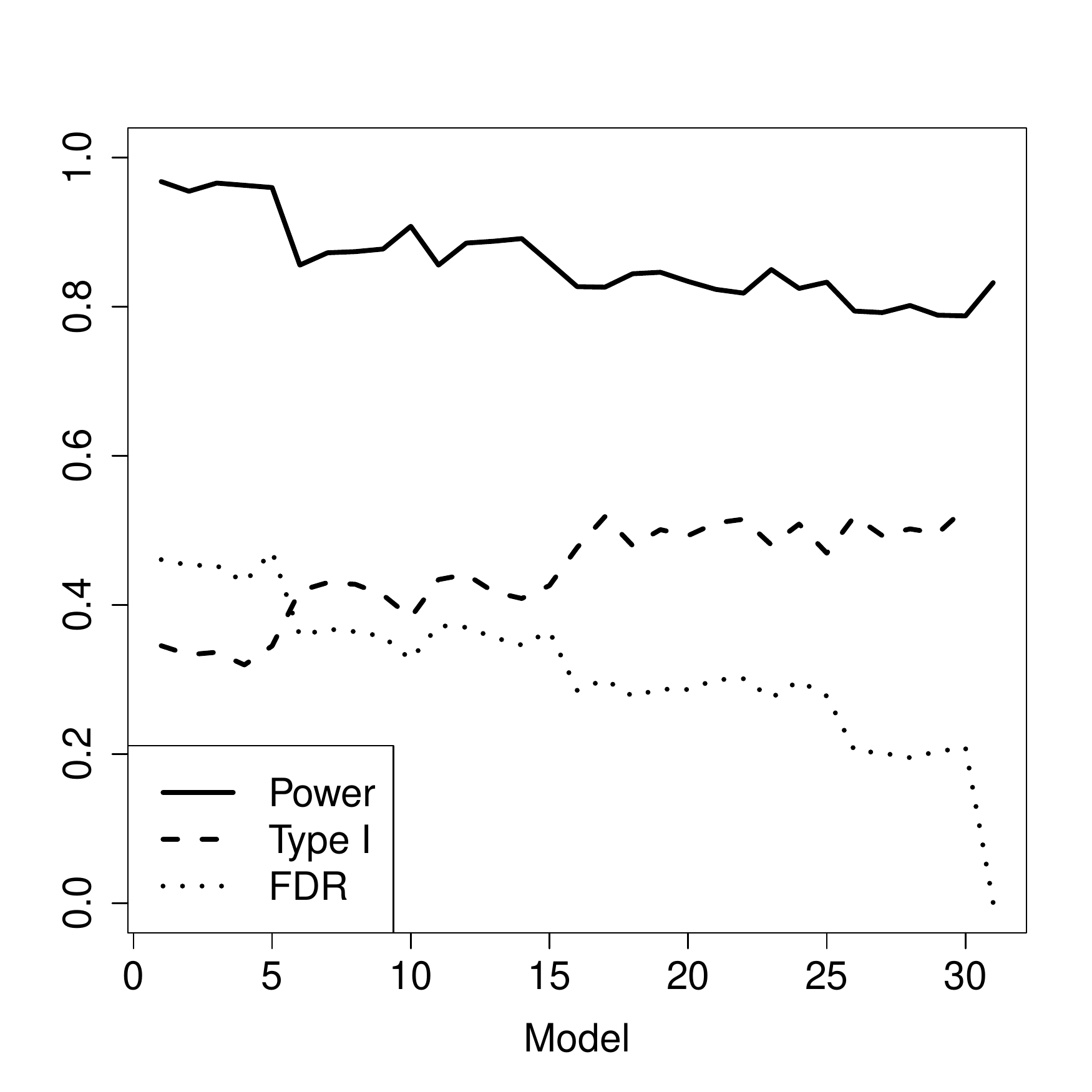} & \includegraphics[scale=\binsimscale]{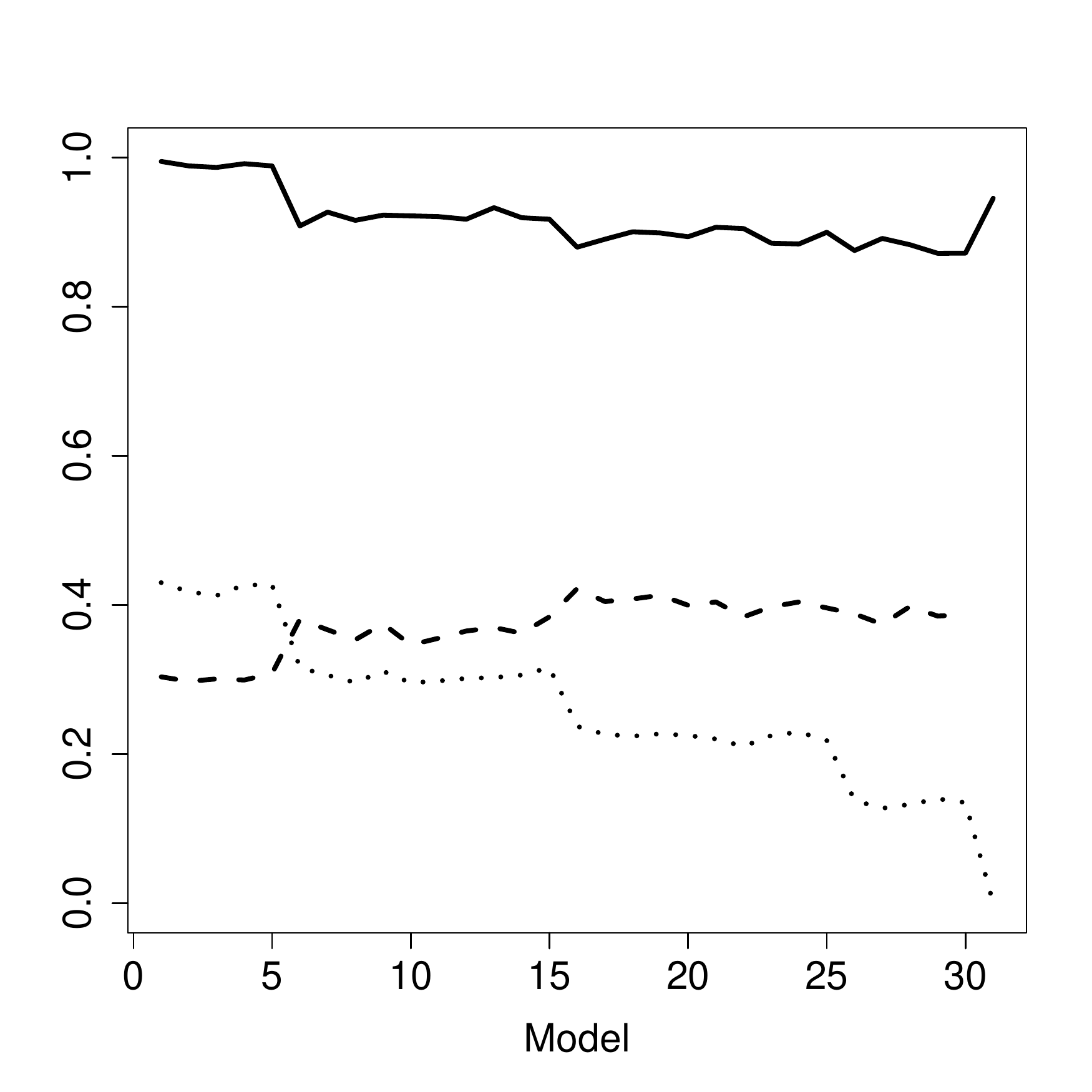} \\
$N = 80$ & $N = 100$ \\[-2ex]
\includegraphics[scale=\binsimscale]{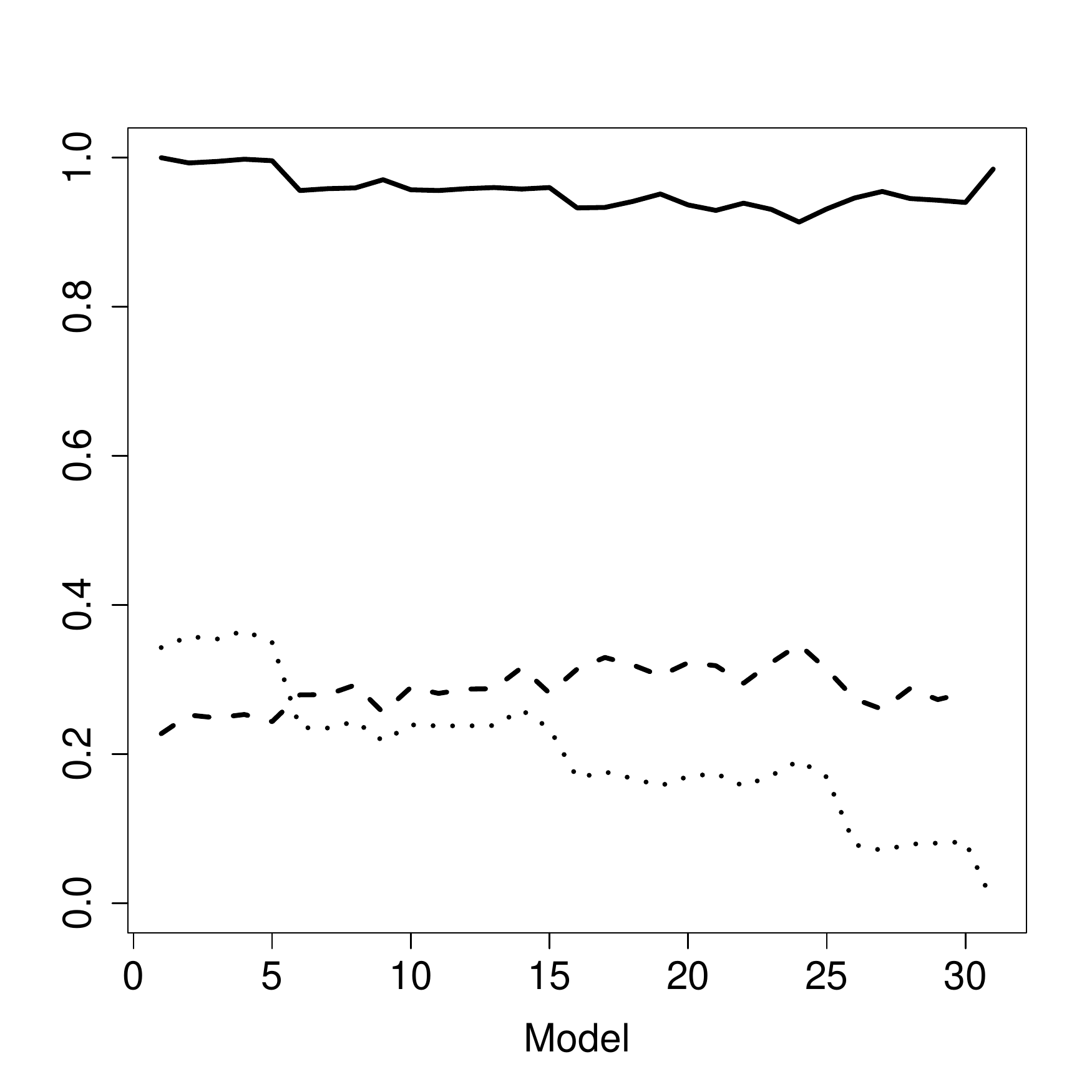} & \includegraphics[scale=\binsimscale]{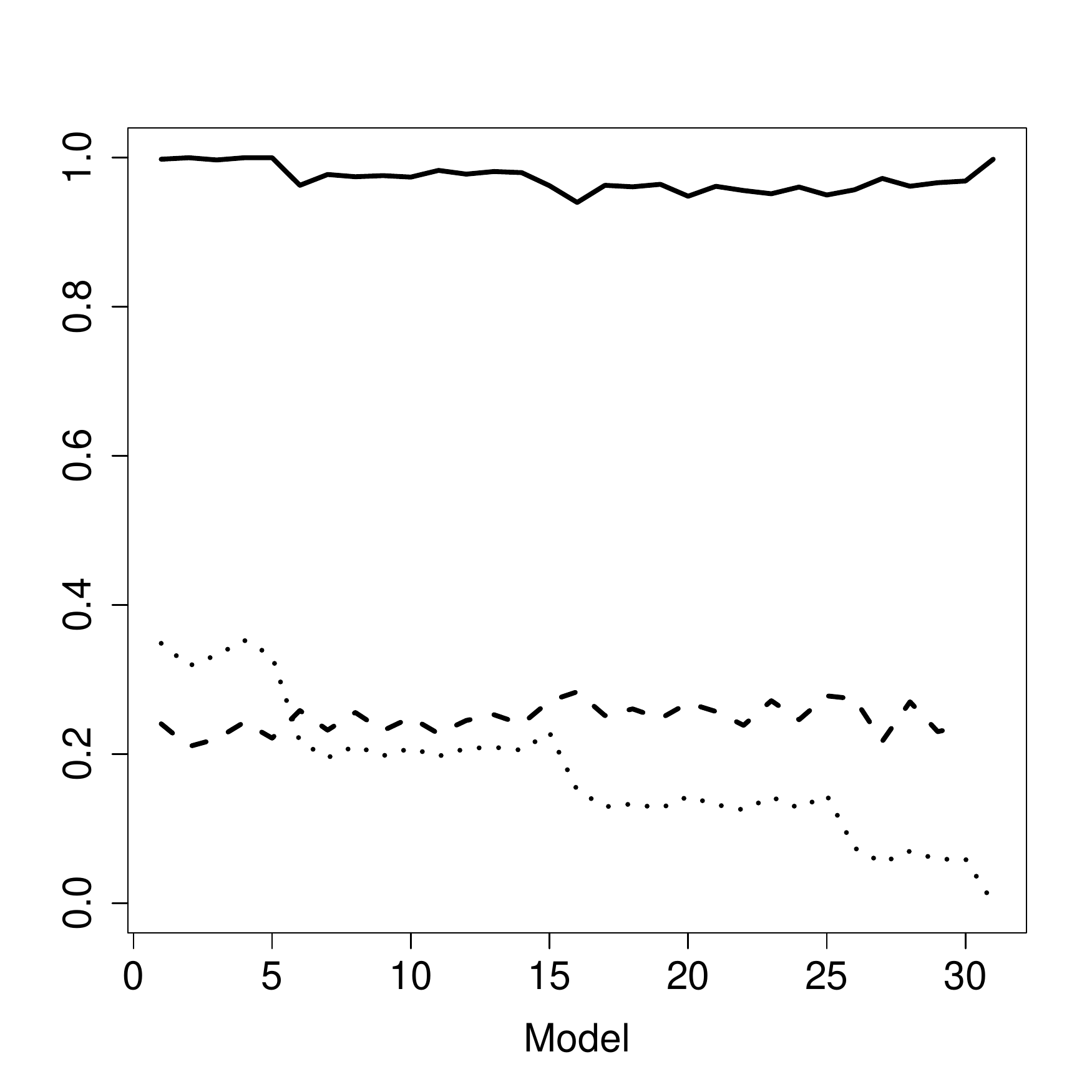}
\end{tabular}
\end{center}
\caption{\label{fig:binsim3}Average power, type I error rate and FDR for logistic regression with $\kappa = 3$ and $N=30,50,80,100$ runs.}
\end{figure}

For all three prior distributions, high power is achieved for all experiment sizes: greater than 80\% for $N=30, 50$, and greater than 90\% for $N=80, 100$. Generally, there is a slight downward trend in power as the size of the true model increases. With the exception of the model including all five variables, a similar trend was observed for $D$-efficiency. The type I error rate is, unsurprisingly, an increasing function of the number of variables in the data-generating (true) model, as there are fewer inactive variables (smaller denominator) for larger true models. The maximum type I error rate of about 0.5 occurs for those true models involving four variables and corresponds to identifying, on average, less than one additional active variable. In contrast, FDR is a decreasing function of true model size, as again there are fewer inactive variables for larger true models. The maximum FDR of approximately 0.5 occurs when $N=30$ for true models that contain a single variable, and corresponds to identifying one fewer additional active variable on average. 

\begin{figure}
\begin{center}
\begin{tabular}{cc}
$N = 30$ & $N = 50$ \\[-2ex]
\includegraphics[scale=\binsimscale]{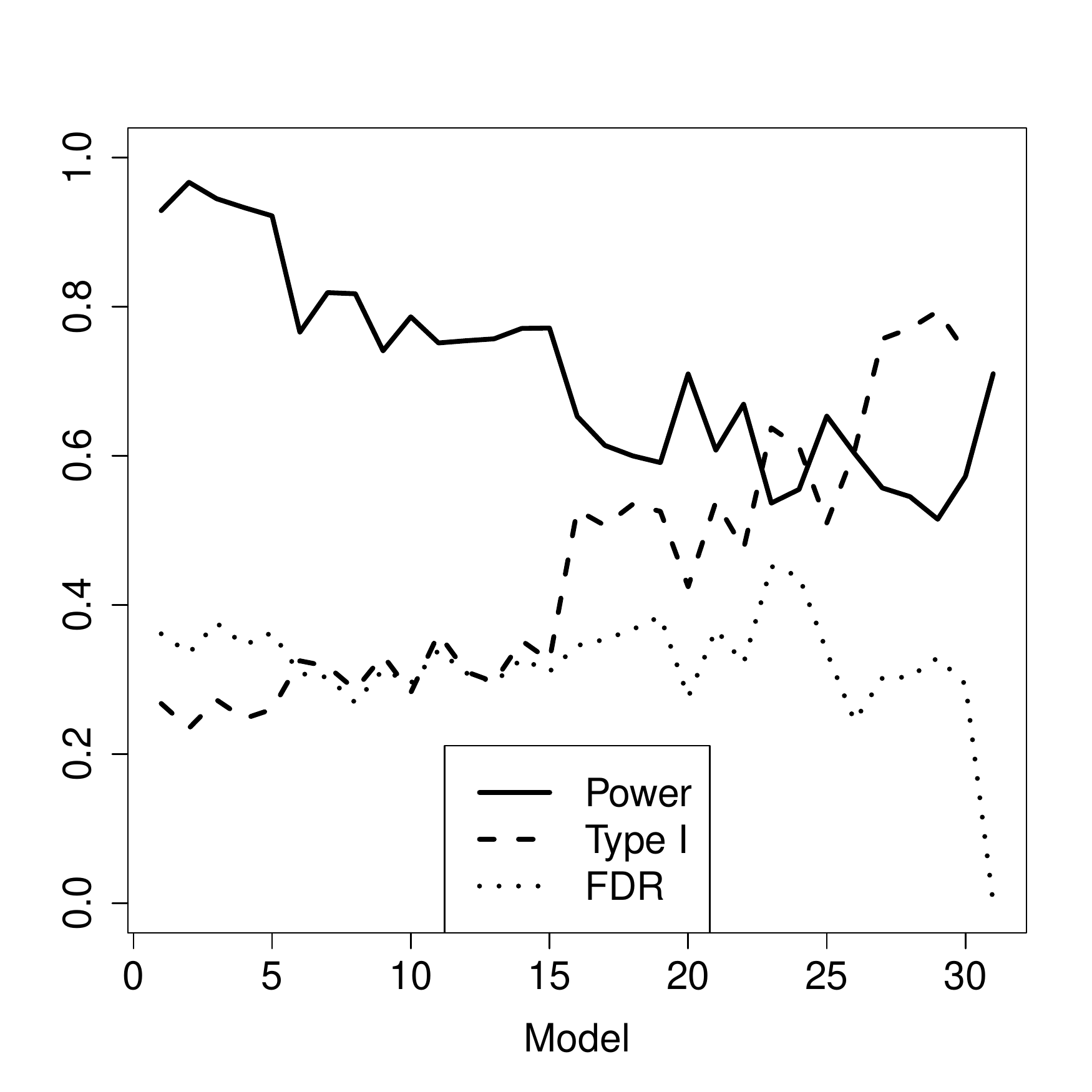} & \includegraphics[scale=\binsimscale]{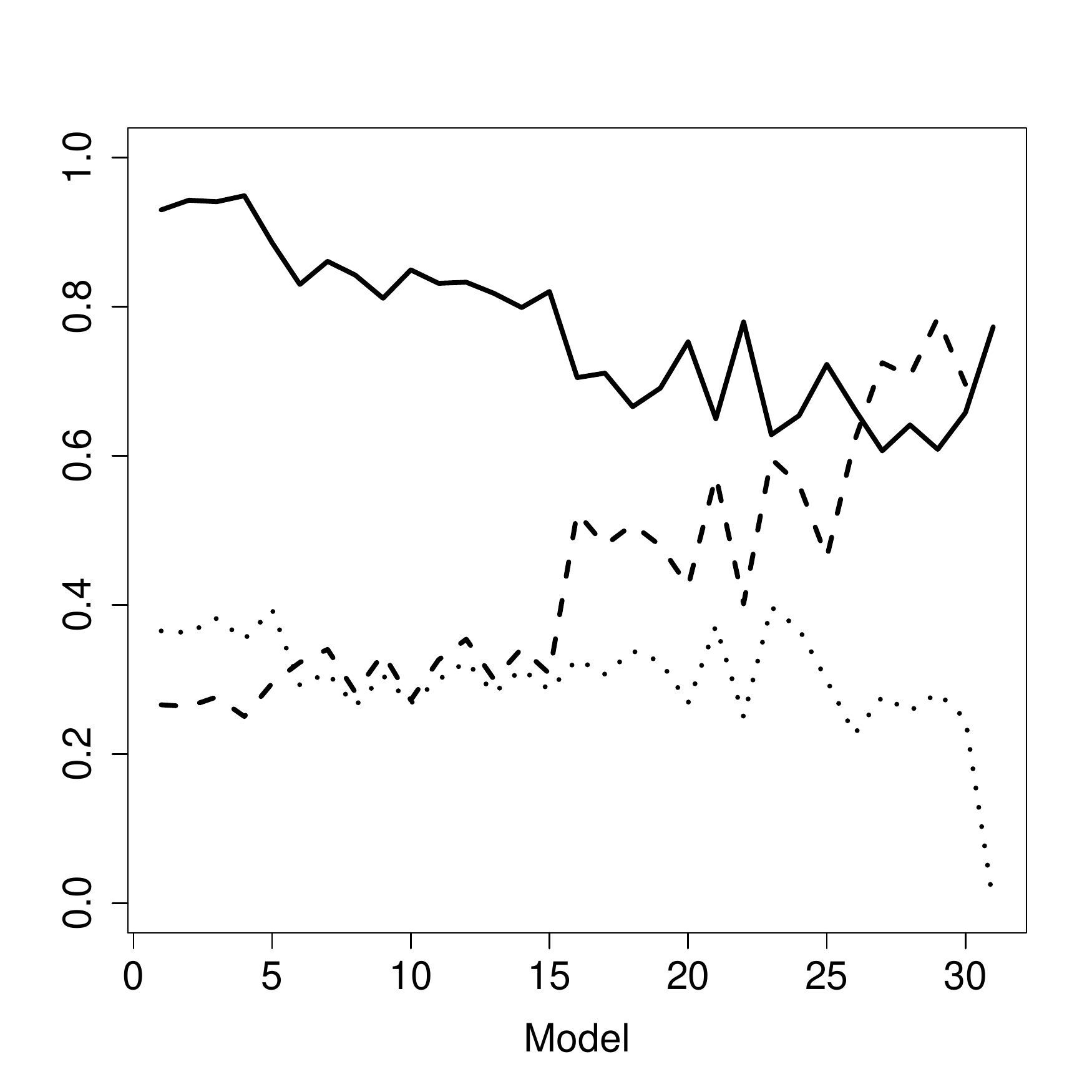} \\
$N = 80$ & $N = 100$ \\[-2ex]
\includegraphics[scale=\binsimscale]{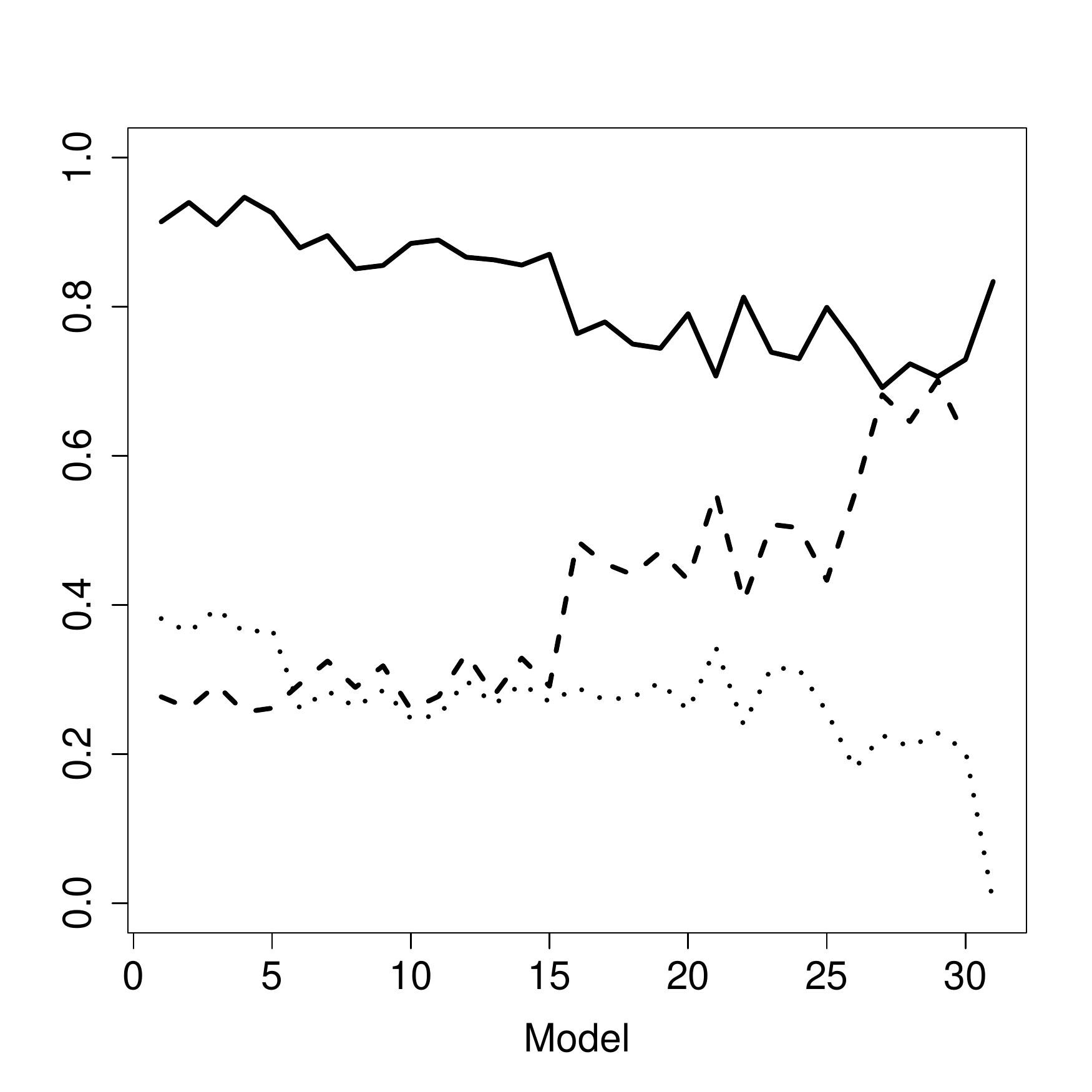} & \includegraphics[scale=\binsimscale]{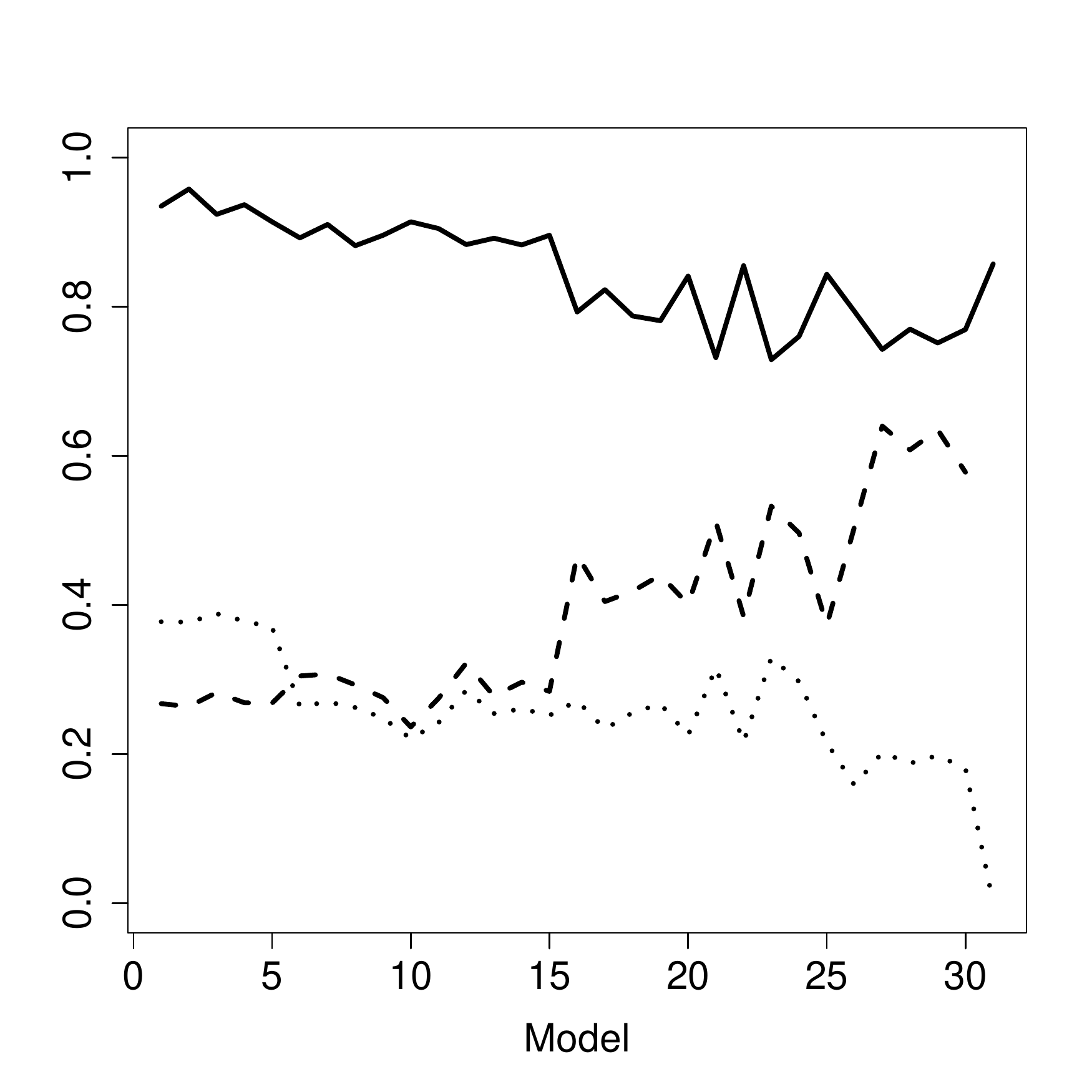}
\end{tabular}
\end{center}
\caption{\label{fig:binsim1local}Average power, type I error rate and FDR for logistic regression with $\kappa = 1$, $N=30,50,80,100$ runs and a locally $D$-optimal design for the maximal model.}
\end{figure}

\begin{figure}
\begin{center}
\begin{tabular}{cc}
$N = 30$ & $N = 50$ \\[-2ex]
\includegraphics[scale=\binsimscale]{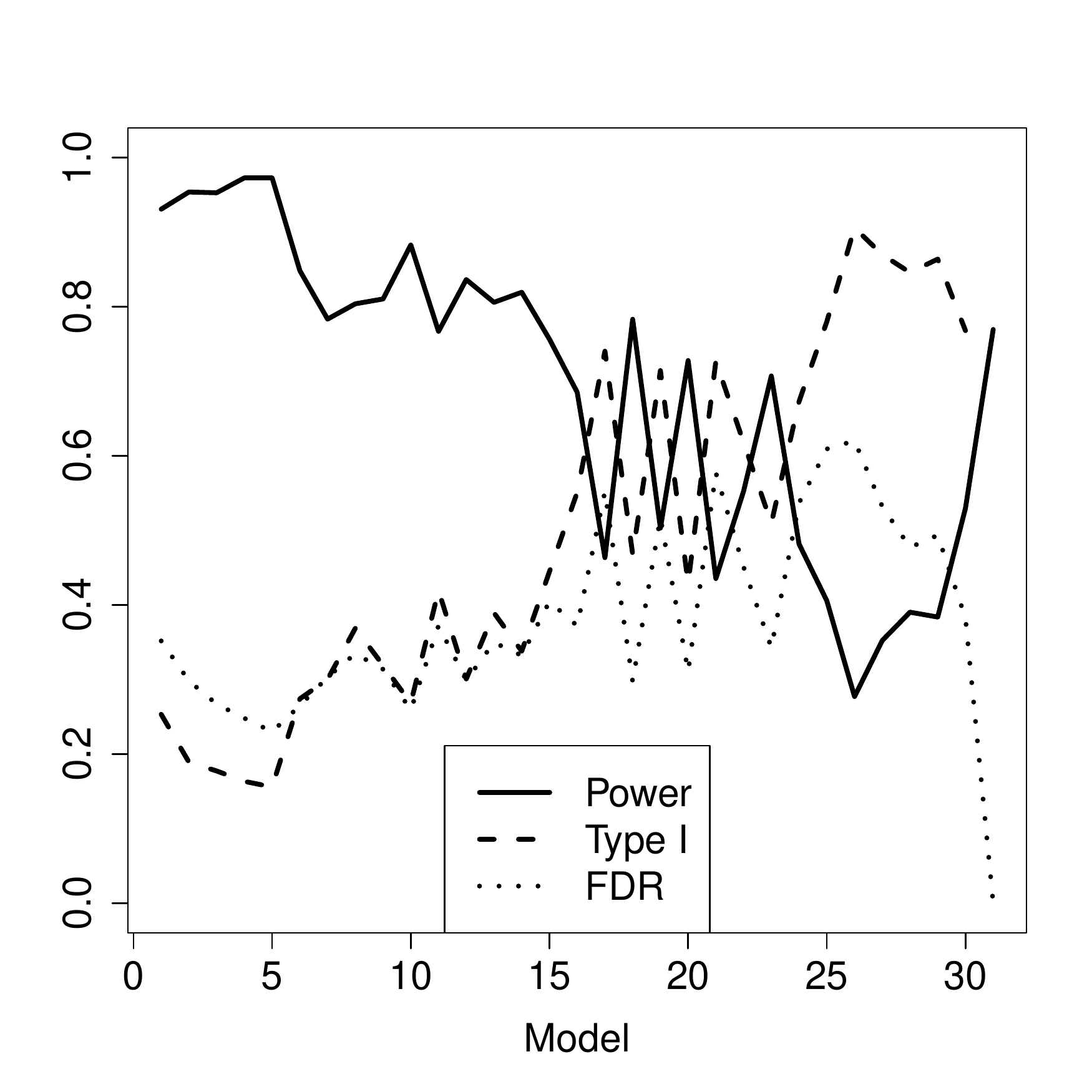} & \includegraphics[scale=\binsimscale]{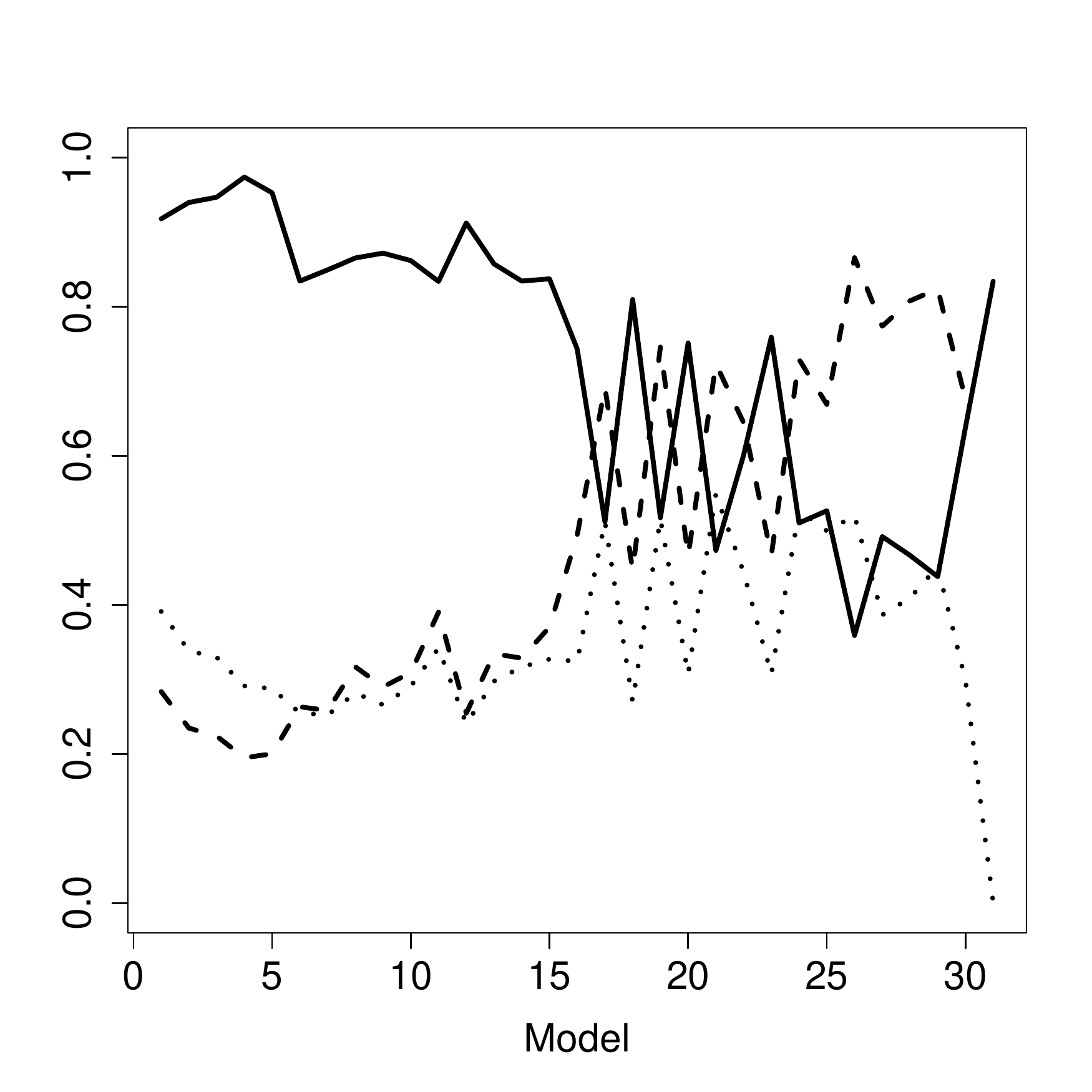} \\
$N = 80$ & $N = 100$ \\[-2ex]
\includegraphics[scale=\binsimscale]{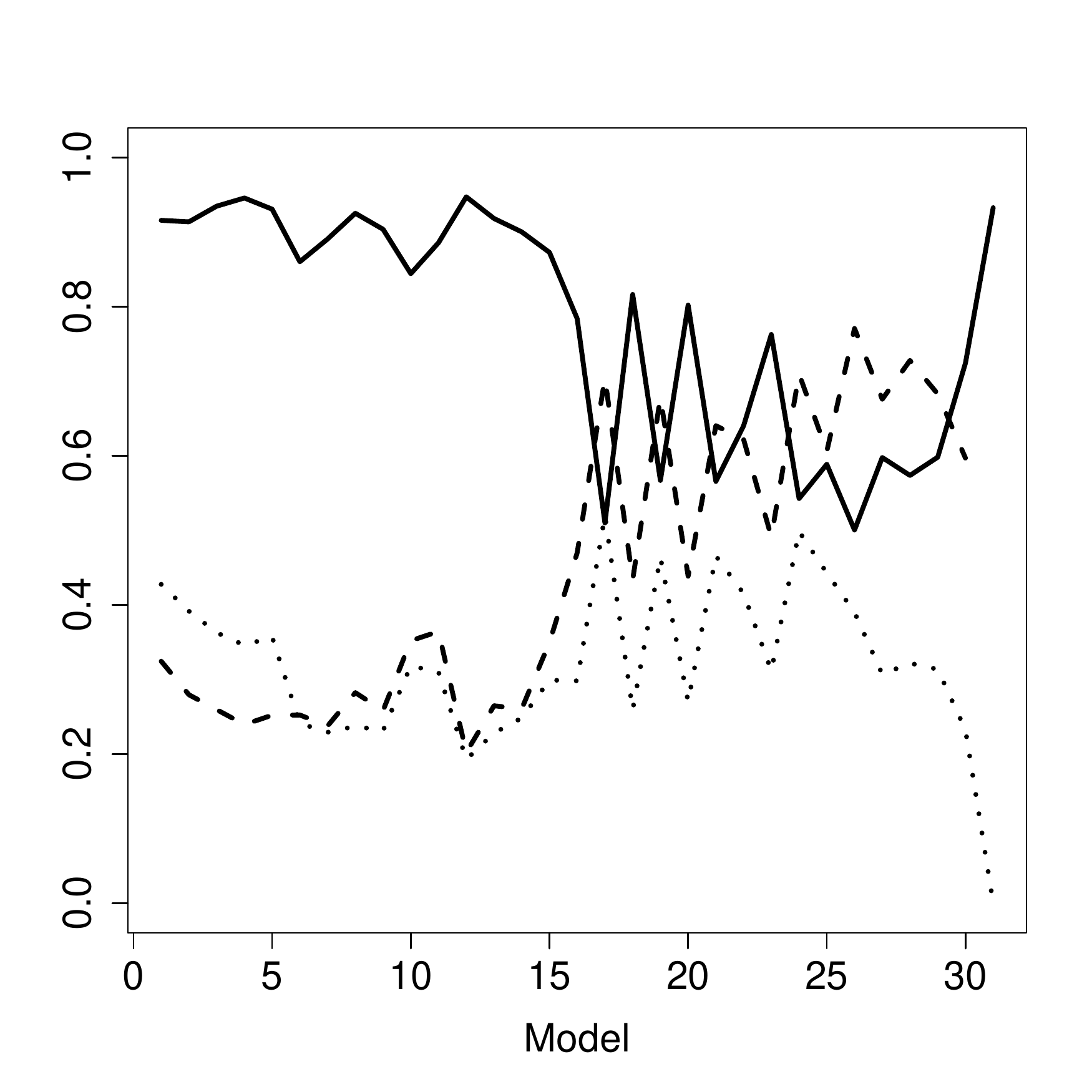} & \includegraphics[scale=\binsimscale]{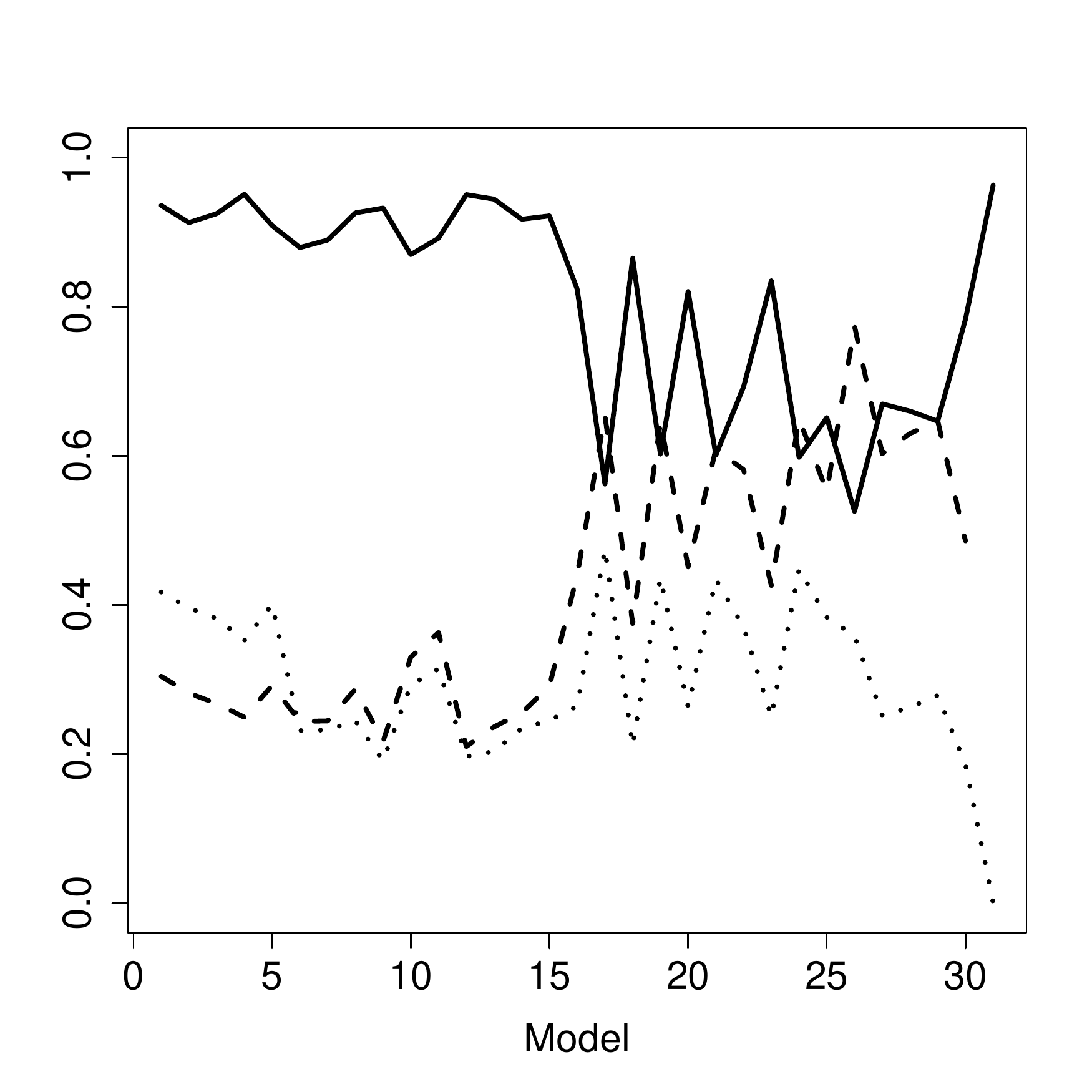}
\end{tabular}
\end{center}
\caption{\label{fig:binsim3local}Average power, type I error rate and FDR for logistic regression with $\kappa = 3$, $N=30,50,80,100$ runs and a locally $D$-optimal design for the maximal linear model.}
\end{figure}

The results are fairly similar for the different prior distributions. The major difference is lower type I error rates when $\kappa = 3$, where there is a greater distinction between the sizes of the model coefficients for active and inactive variables. For comparison, we present results obtained from using the locally $D$-optimal design for model 31 (Figure~\ref{fig:binsim1local} for $\kappa=1$ and Figure~\ref{fig:binsim3local} for $\kappa=3$). In general, the average power is lower and average Type I error and false discovery rates higher for this design than for the Bayesian IC designs. For $\kappa=3$, the locally optimal design has particularly poor performance for some models having three or four variables.

Another obvious comparator for the Bayesian IC designs are locally optimal IC designs, i.e. designs that maximise~\eqref{eq:locoptIC} with parameters $\bbeta_m$ set equal to the mean of prior distribution~\eqref{eq:prior} ($m=1,\ldots,M$). For this example, the Bayesian IC and locally optimal IC designs have almost identical $D$-efficiency distributions and model selection results, with the Bayesian IC designs displaying very slightly less variable $D$-efficiencies when there are $n=6$ support points. These results (which are not shown) further illustrate the generally good performance of IC designs for model selection. We anticipate greater differences between Bayesian and locally optimal IC designs for examples where the prior distribution does not specify a known sign for each model parameter.

\begin{figure}
\begin{center}
\begin{tabular}{cc}
$N = 30$, Model 1 & $N = 100$, Model 1 \\[-2ex]
\includegraphics[scale=\binsimscale]{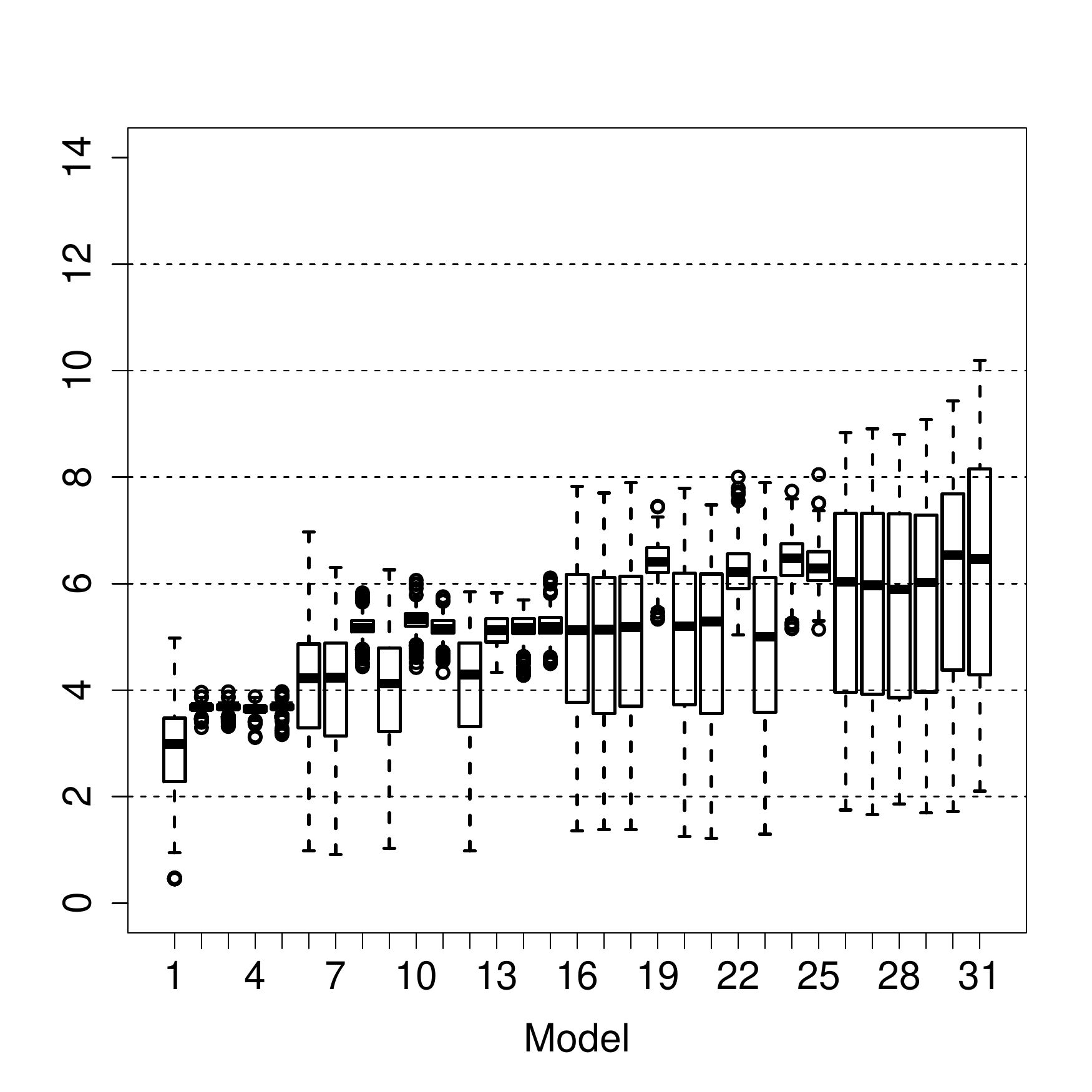} & \includegraphics[scale=\binsimscale]{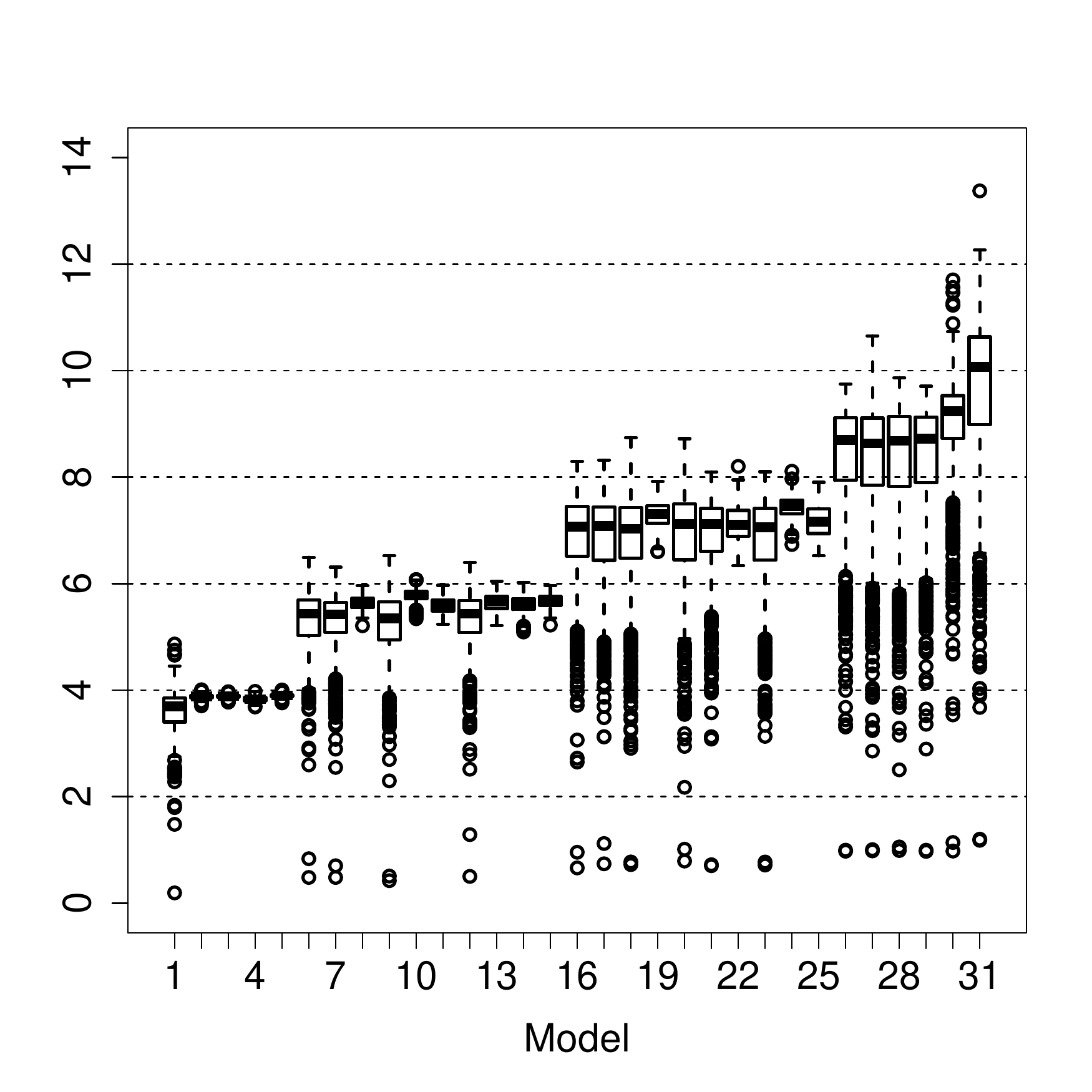} \\
$N = 30$, Model 16 & $N = 100$, model 16 \\[-2ex]
\includegraphics[scale=\binsimscale]{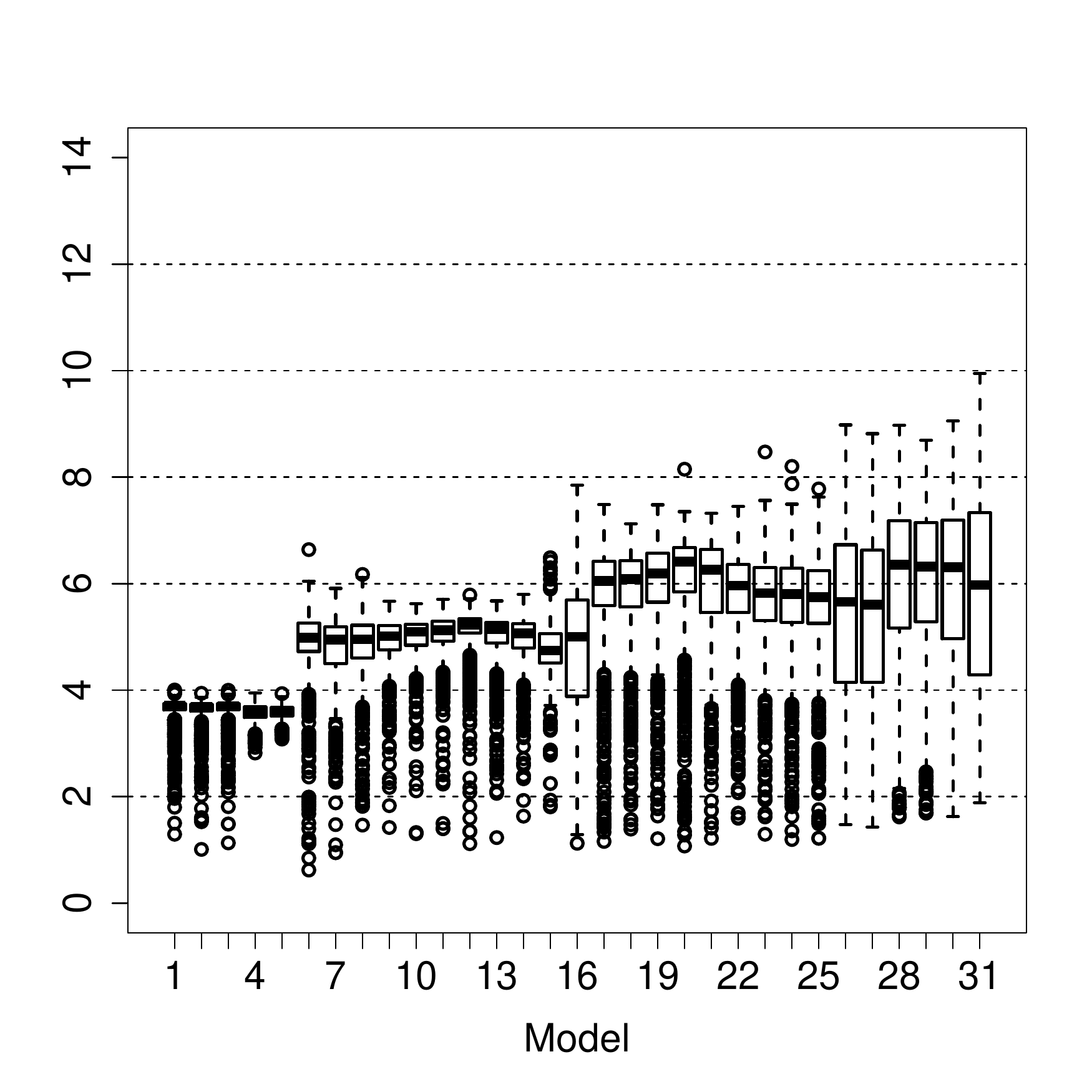} & \includegraphics[scale=\binsimscale]{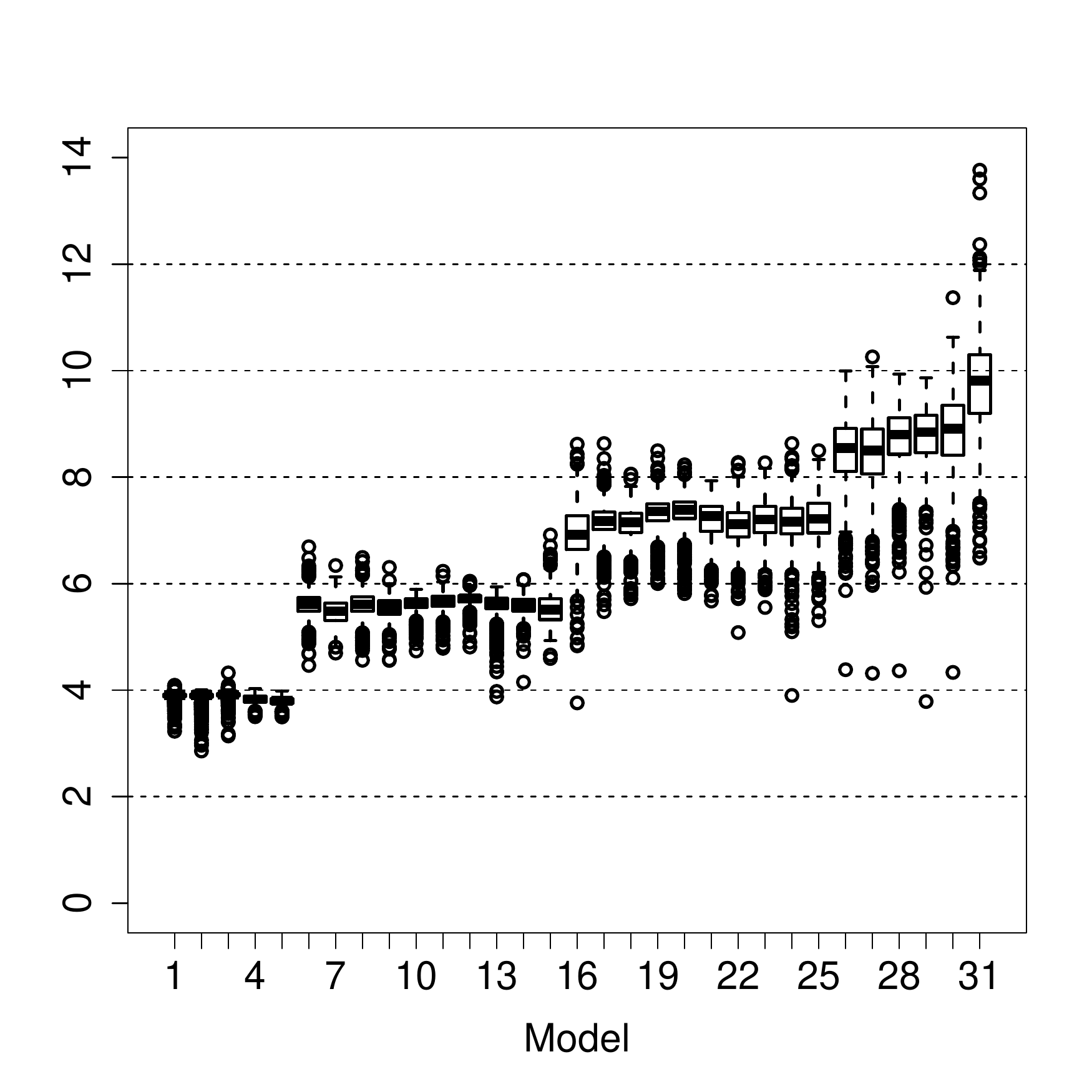}
\end{tabular}
\end{center}
\caption{\label{fig:pen}Boxplots of the GIC penalty from~\eqref{eq:gic} with $\kappa = 1$ and data generated from models 1 (variable 1 only) and 16 (variables 1 and 2 only).}
\end{figure}

A key determinant of the model selection findings is the size of the GIC penalty term in~\eqref{eq:gic}. Our numerical studies have shown that this depends not only on the size, $p_m$, of the model but also on the estimated model parameters and the goodness of fit, with better-fitting models having a smaller penalty. Figure~\ref{fig:pen} shows the distributions of the penalties obtained when model 1 (variable 1) and model 16 (variables 1 and 2) are true for $N=30, 100$ and $\kappa=1$. In general, the penalty is somewhat less than $2p_m$, although it increases with $N$ and hence does not penalise larger models to the same degree as AIC. Models that include the correct variables have smaller penalty than other models. Further research on the use of this penalty is needed.

\section{Design and model selection for Poisson response and log-linear regression}\label{sec:poisson}

To investigate the performance of the methodology for log-linear regression, simulation studies were performed for two examples. Both assume the log link, $g(\mu) = \eta$, and linear predictors of the form~\eqref{eq:IClinpred}. In the first example, there are again $q = 5$ variables that may affect the response (31 possible models) and prior distribution~\eqref{eq:prior} is assumed. In the second example, there are $q=10$ variables but, in line with factor sparsity \citep{BM1986}, we consider only linear predictors including at most three active variables (175 possible models). Prior distributions for $\beta_{im}$ are given by~\eqref{eq:prior} for $i=1,\ldots,5$ and for the remaining parameters by

\begin{equation*}\label{eq:prior2}
\beta_{im} \sim \left\{
\begin{array}{ll}
\mbox{Uniform}(\kappa, 5) & \mbox{for } i = 6,8,9 \mbox{ and } I(i, m) = 1\,, \\
\mbox{Uniform}(-5, -\kappa) & \mbox{for } i = 7,10 \mbox{ and } I(i, m) = 1\,. \\
\end{array}
\right.
\end{equation*}
Again, $\beta_{0m}=0$, $\beta_{im}=0$ if $I(i,m)=0$ and $\kappa=1,2,3$.

\subsection{Minimally-supported designs}\label{sec:poisic}

\newcommand{\poissdeffscale}{.3}
\begin{figure}
\begin{center}
\begin{tabular}{cc}
(a) & (d) \\[-2ex] 
\includegraphics[scale=\poissdeffscale]{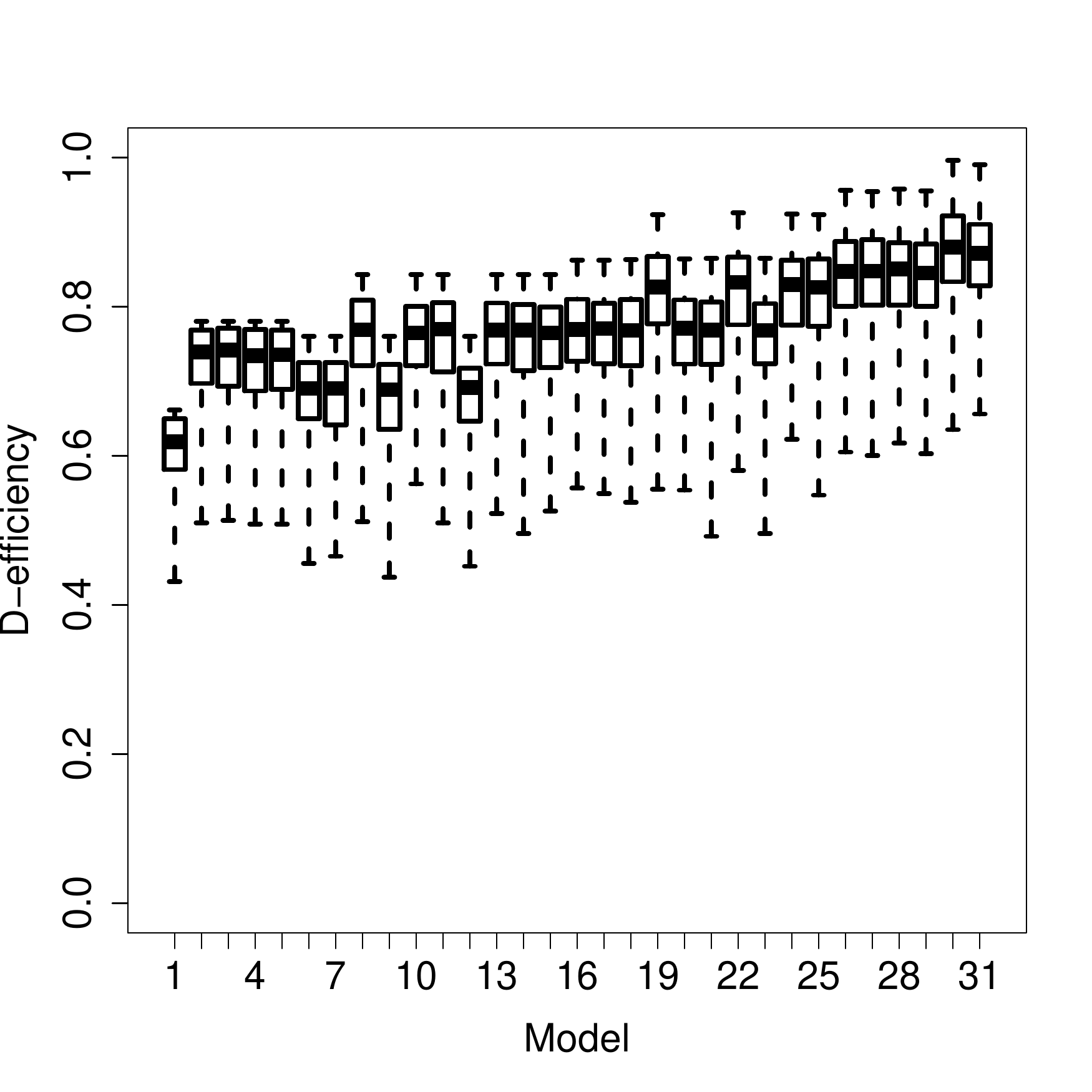} & \includegraphics[scale=\poissdeffscale]{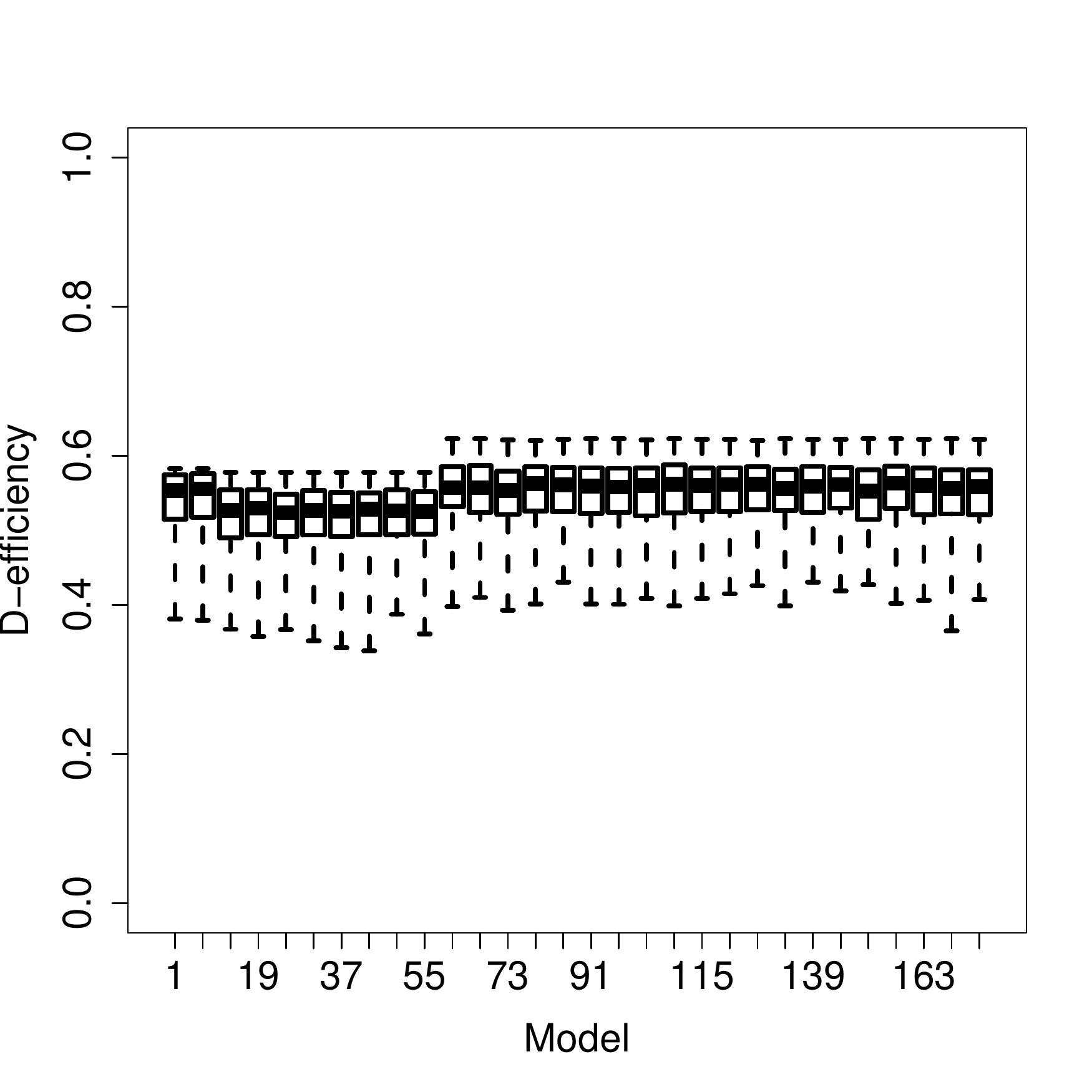} \\  
(b) & (e) \\[-2ex]
\includegraphics[scale=\poissdeffscale]{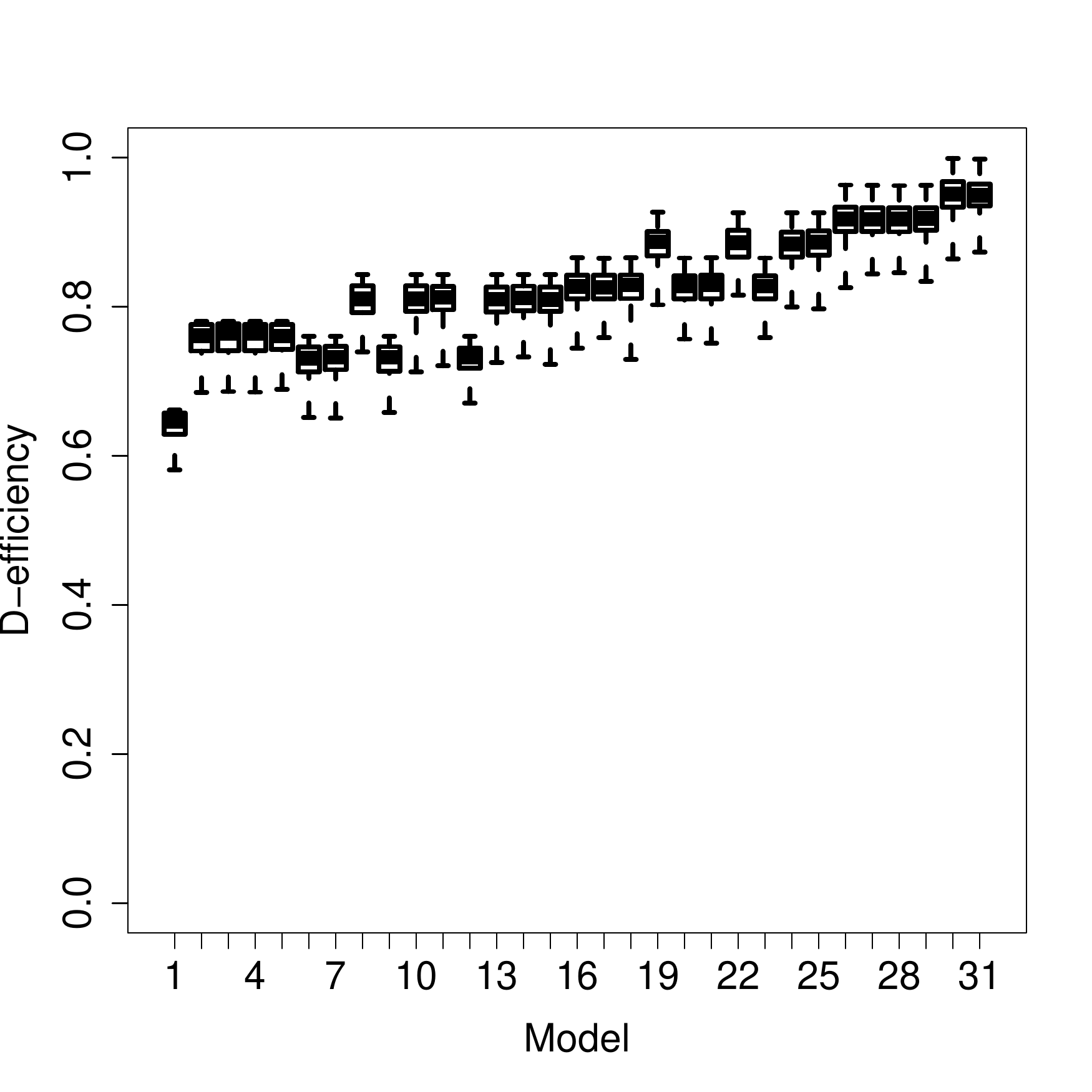} & \includegraphics[scale=\poissdeffscale]{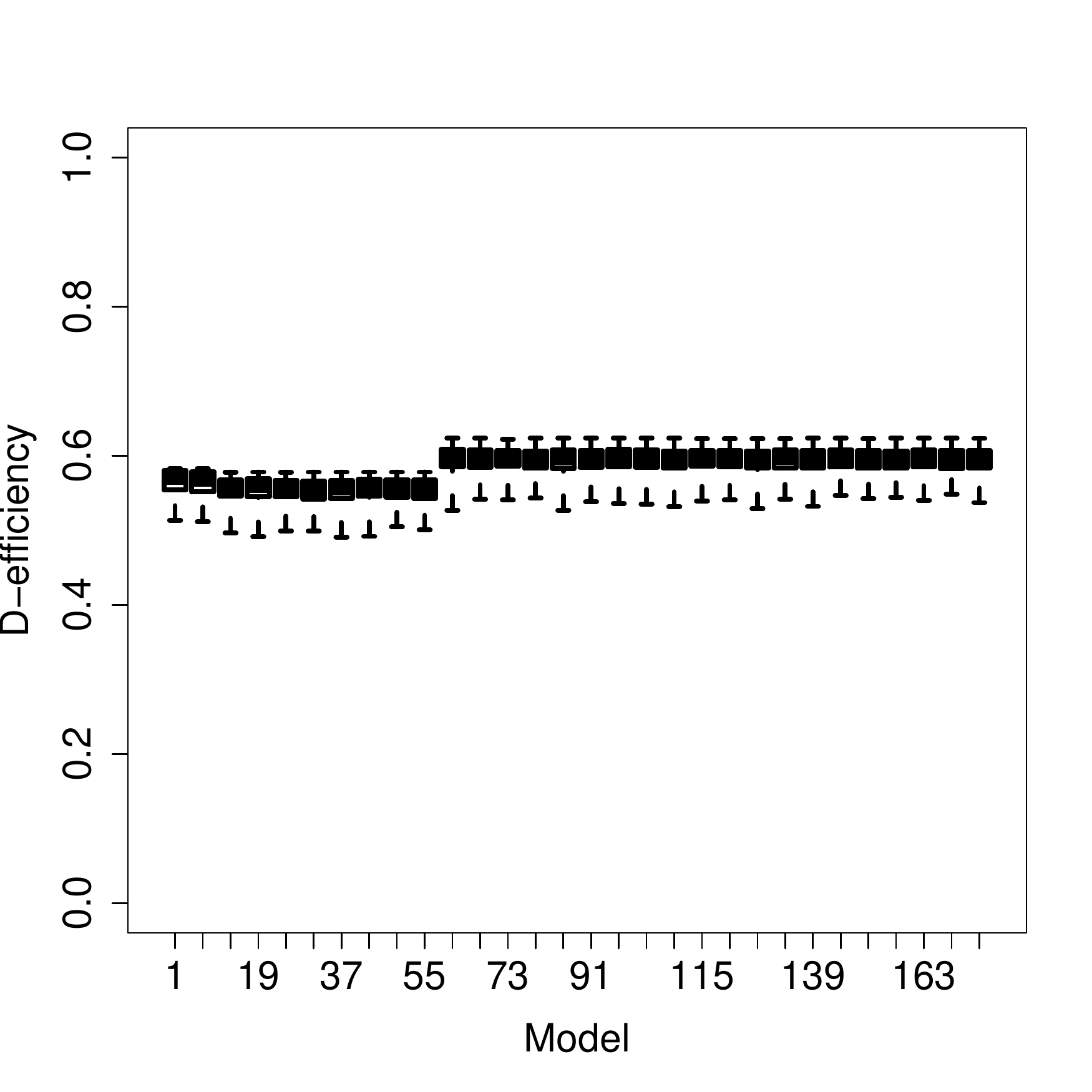} \\
(c) & (f) \\[-2ex]
\includegraphics[scale=\poissdeffscale]{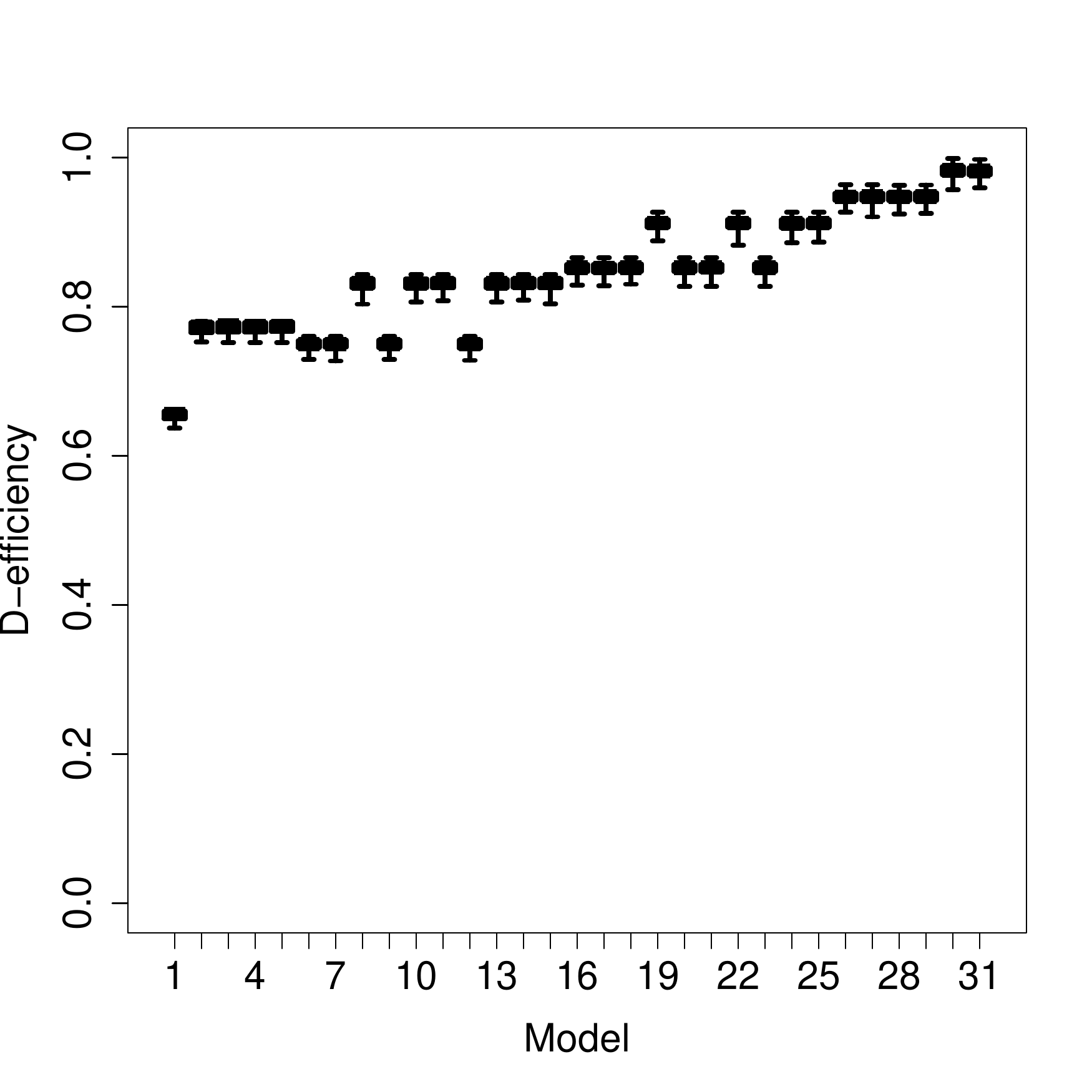} & \includegraphics[scale=\poissdeffscale]{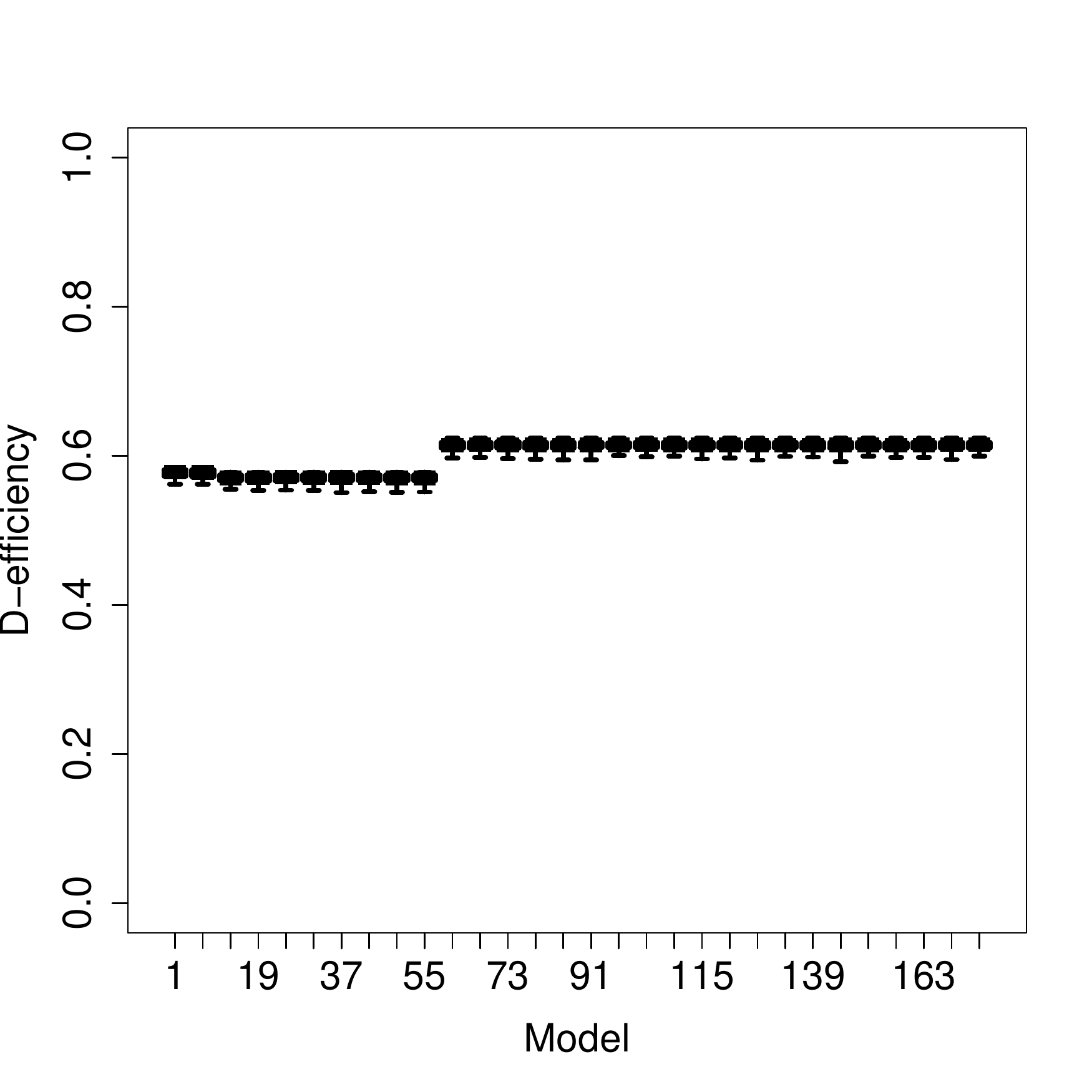} \\
\end{tabular}
\end{center}
\caption{\label{fig:poissdeff}Boxplots of $D$-efficiencies for robust designs for log-linear regression: five variables (a) $\kappa = 1$, (b) $\kappa = 2$ and (c) $\kappa = 3$; 10 variables (d) $\kappa = 1$, (e) $\kappa = 2$ and (f) $\kappa = 3$. For the 10 variable case, only results for every sixth model are displayed.}
\end{figure}

\begin{figure}
\begin{center}
\begin{tabular}{cc}
(a) & (d) \\[-2ex] 
\includegraphics[scale=\poissdeffscale]{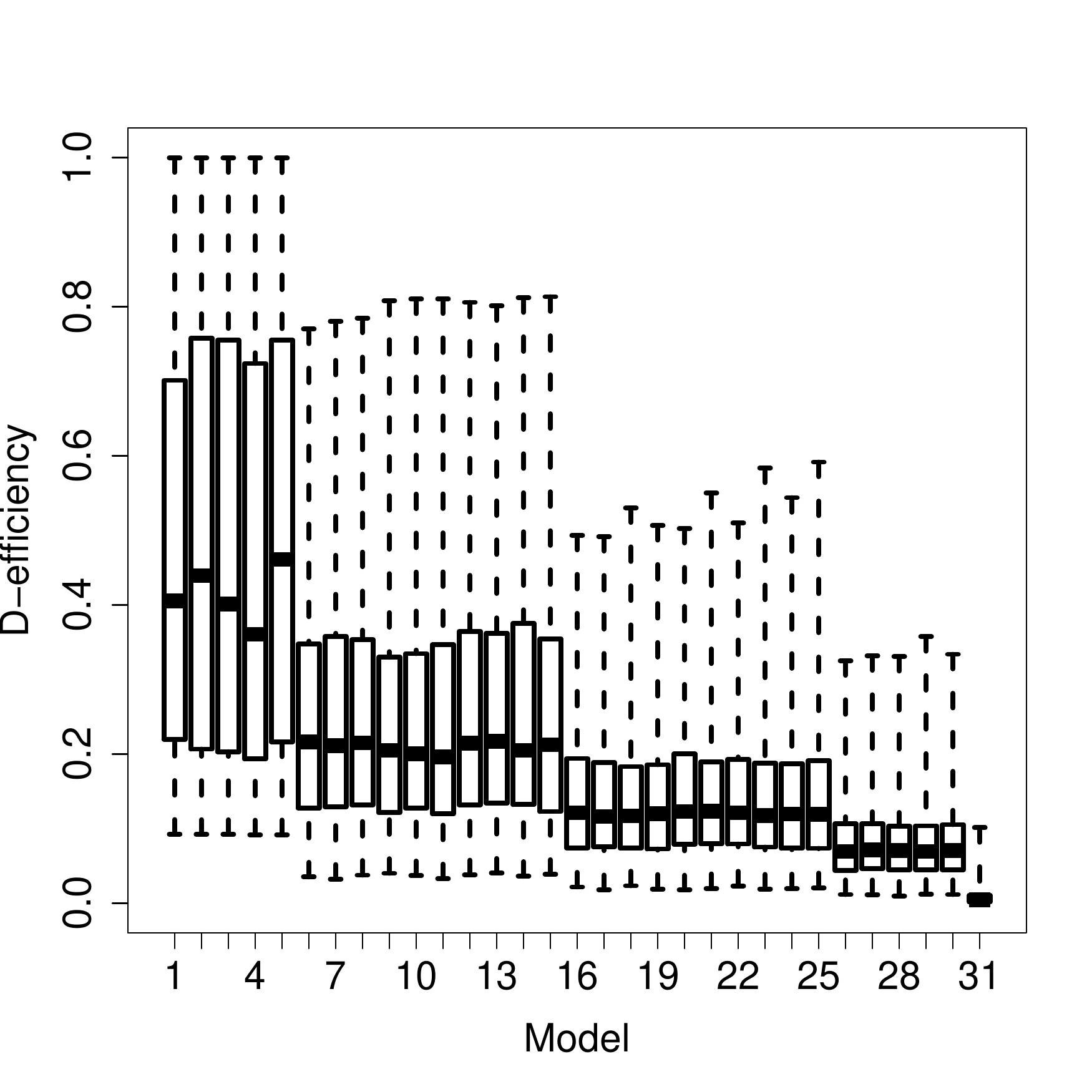} & \includegraphics[scale=\poissdeffscale]{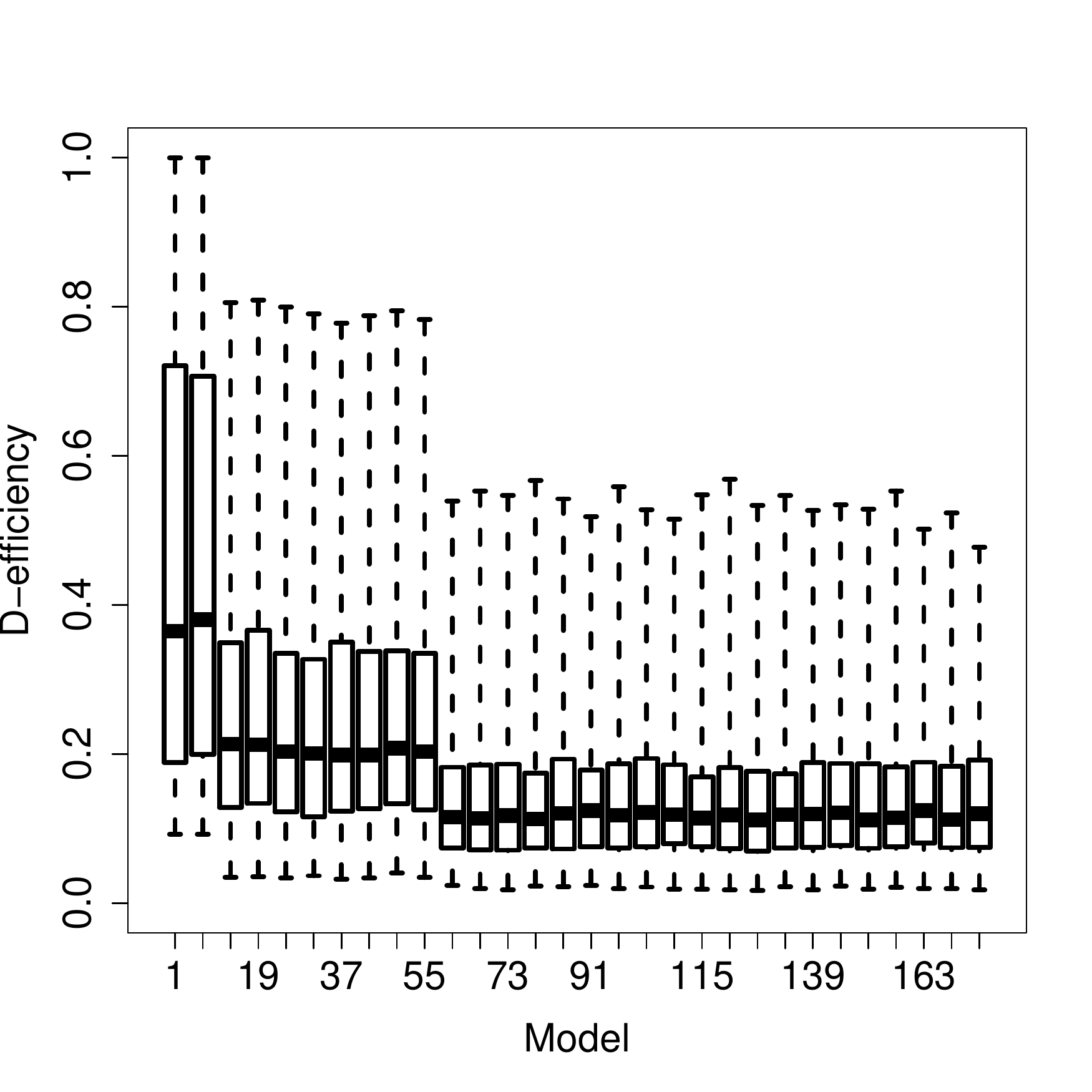} \\  
(b) & (e) \\[-2ex]
\includegraphics[scale=\poissdeffscale]{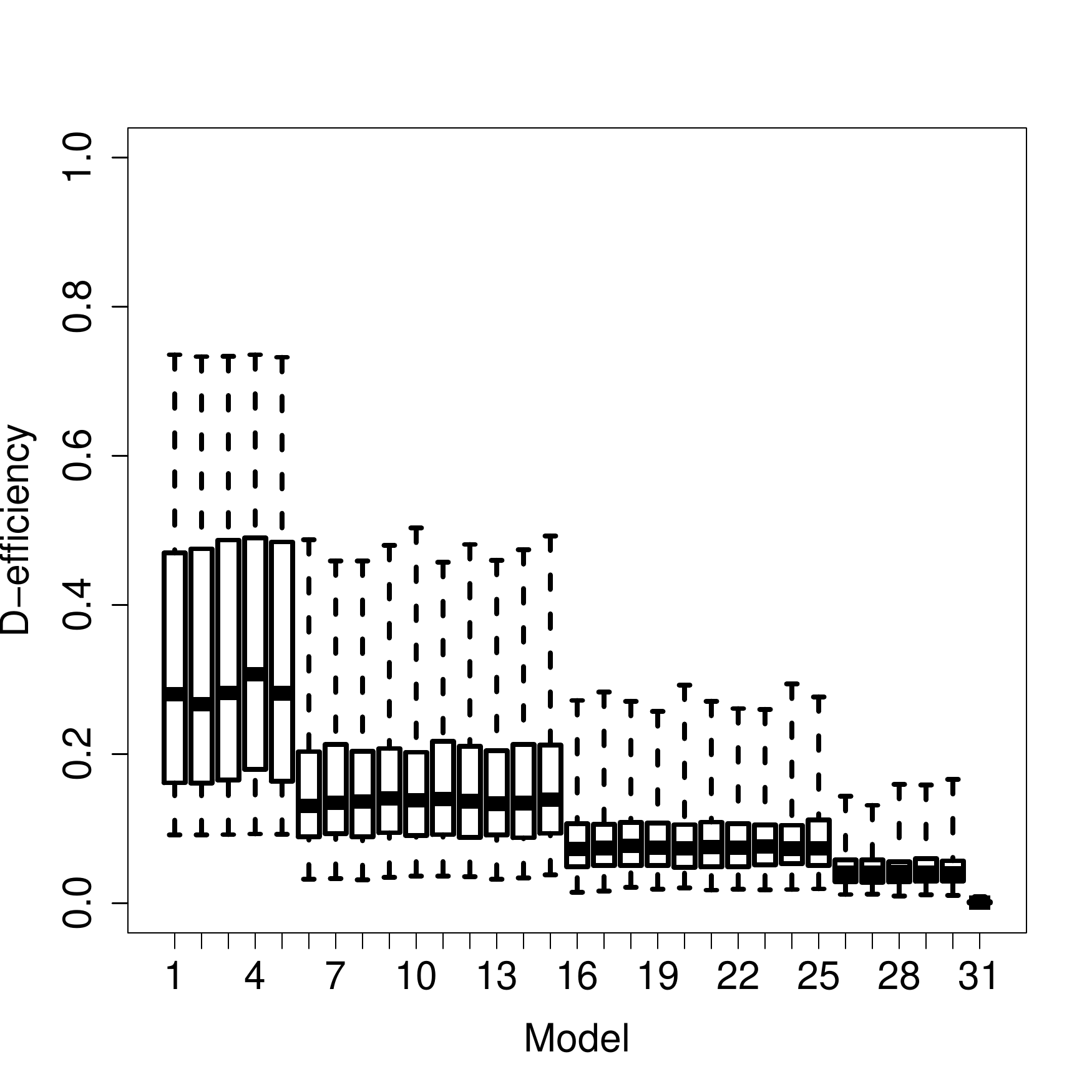} & \includegraphics[scale=\poissdeffscale]{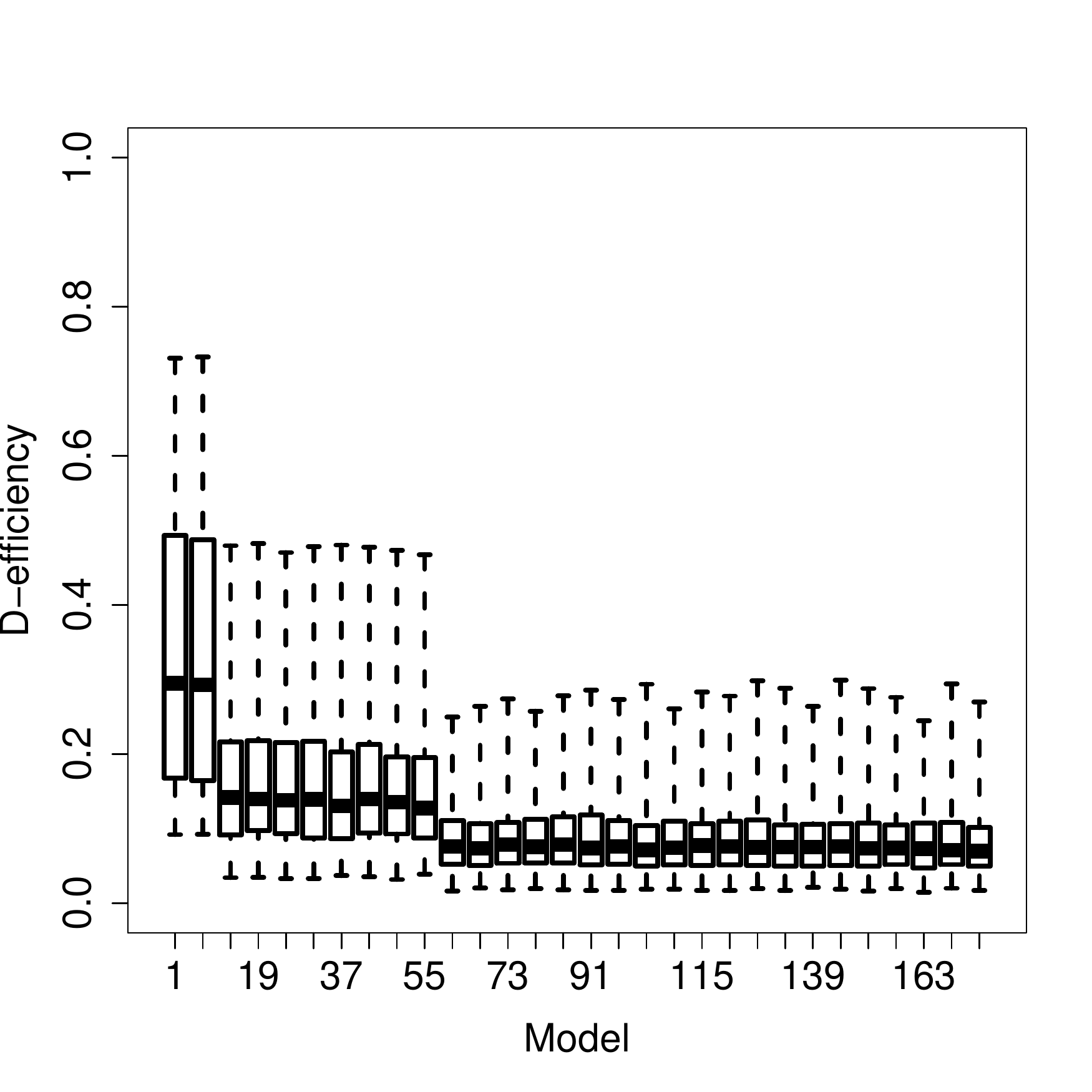} \\
(c) & (f) \\[-2ex]
\includegraphics[scale=\poissdeffscale]{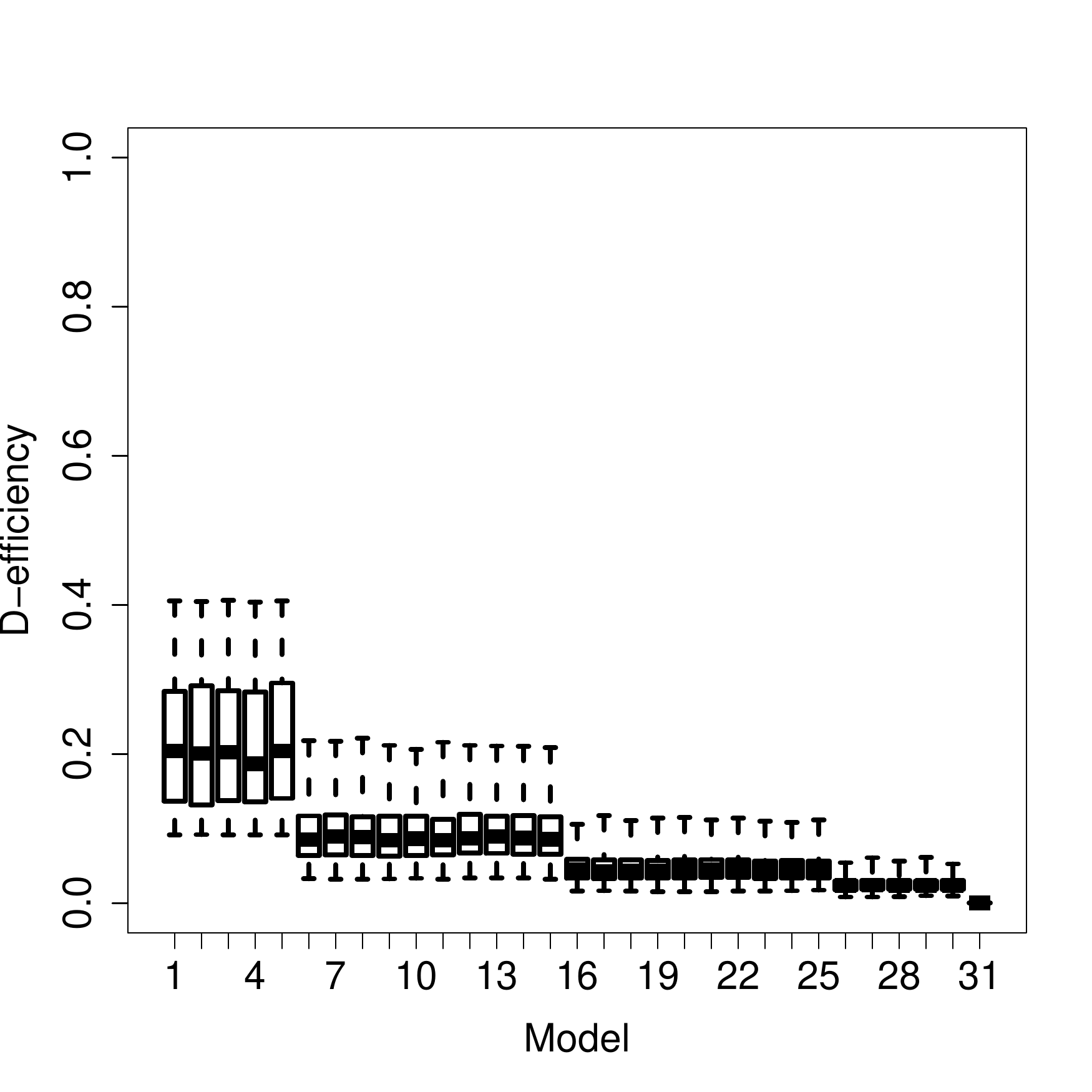} & \includegraphics[scale=\poissdeffscale]{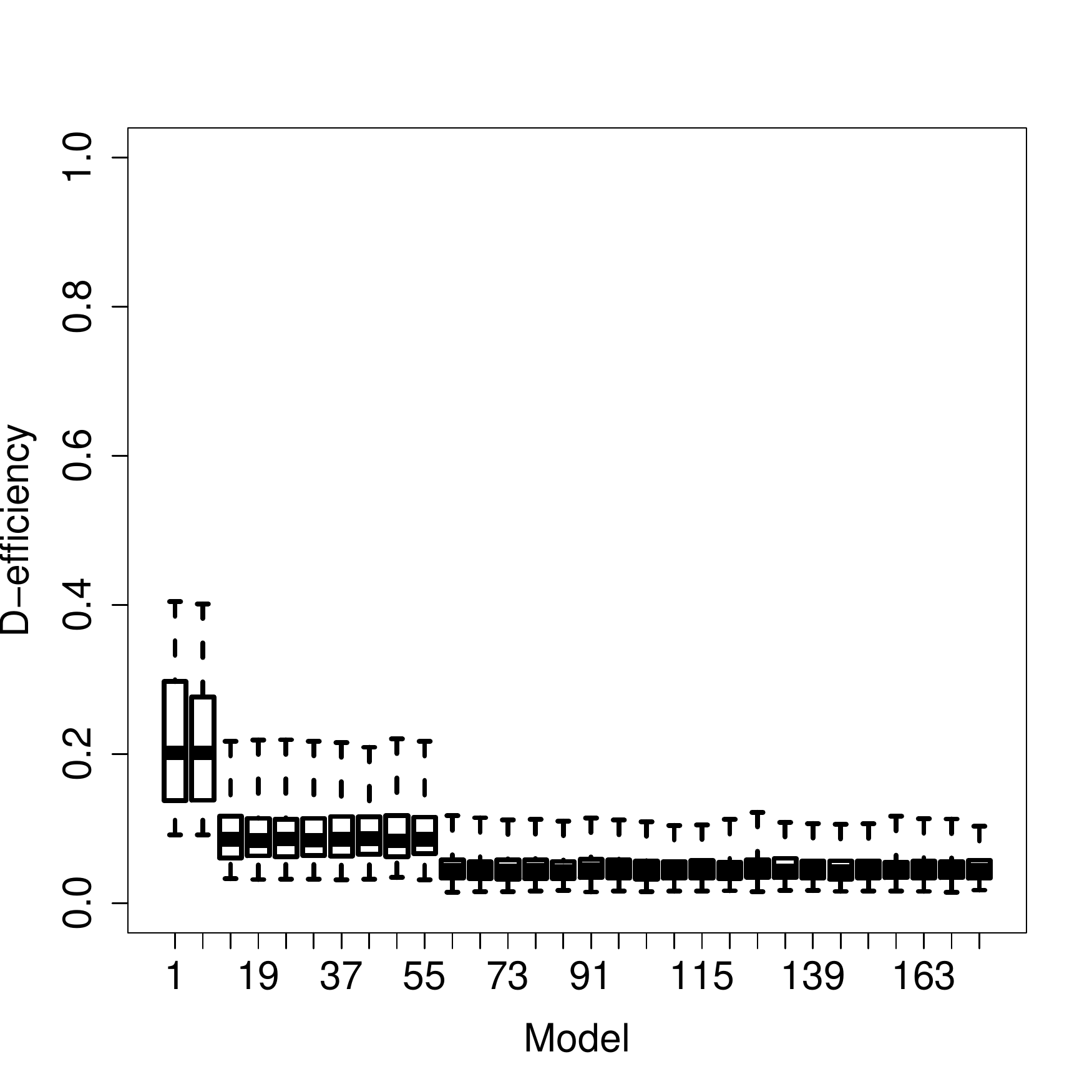} \\
\end{tabular}
\end{center}
\caption{\label{fig:poissdefflocal}Boxplots of $D$-efficiencies for fractional factorial designs for log-linear regression: five variables (a) $\kappa = 1$, (b) $\kappa = 2$ and (c) $\kappa = 3$; 10 variables (d) $\kappa = 1$, (e) $\kappa = 2$ and (f) $\kappa = 3$. For the 10 variable case, only results for every sixth model are displayed.}
\end{figure}

We restrict attention to designs that are minimally supported with respect to the maximal model, that is, where the number, $n$, of distinct support points is $q+1$. For this class of designs, \citet{RWLE2009} and \citet{ME2012} presented analytical design construction methods. \citet{AW2015} showed that for these designs with $-1\le x_{ij} \le 1$ and $E(\beta_{im})\ge 1$, for any $m=1,\ldots,M$,
$$
\int _{\mathcal{B}_M} \log\mbox{det}\left\{ X_m^\mathrm{T} W_mX_m\right\}\pi_m(\bbeta_m)\,\mathrm{d}\bbeta_m = \log\mbox{det}\left\{X_m^\mathrm{T} W_m^\star X_m\right\}\,,
$$
where $W_m^\star = \mathrm{diag}\{e^{\eta_{jm}^\star}\}$ with $\eta_{jm}^\star = f(\bx_j)^{\mathrm{T}}\bbeta_m^\star$ and $\bbeta_m^\star = E(\bbeta_m)$, the prior expectation of $\bbeta_m$. Hence, numerical integration is no longer required for design evaluation. To exploit the available theory, we find designs that maximise
\begin{equation}\label{eq:PoissIC}
\Phi_D(\xi) = \log\mbox{det}\left\{X_M^\mathrm{T} W_M^\star X_M\right\}\,.
\end{equation}

Maximisation of~\eqref{eq:PoissIC} defines a (pseudo-) Bayesian $D$-optimality criterion for the maximal model. A heuristic justification for using this criterion to find model-robust designs was given by \citet{ME2012} who pointed out that, assuming common prior distributions, the levels included for each variable in the minimally-supported Bayesian $D$-optimal design for each individual model $m$ are the same. Only the numbers of replications of each variable value differ between the designs. Hence the sub-designs defined as projections of the minimally-supported design for the maximal model into a subset of the variables will contain the same values of the variables as a minimally-supported optimal design for that subset of variables but with different replication. Typically, designs defined in this way display less balance in the variable levels than the $D$-optimal designs for the sub-models. The advantage of maximising~\eqref{eq:PoissIC} is that no numerical optimisation is required for design selection, and hence large examples (e.g. 10 variables) can be investigated.  

We replicate minimally-supported Bayesian $D$-optimal designs that maximise~\eqref{eq:PoissIC} to obtain designs with $N=16$ runs for five variables and $N=32$ runs for 10 variables. Figure~\ref{fig:poissdeff} summarises, for each model, the $D$-efficiencies~\eqref{eq:Deff} for the five variable and 10 variable designs, calculated as described in Section~\ref{sec:binic} except that, in step (ii), the locally $D$-optimal designs obtained from the Theorem of \citet{RWLE2009} are used. In general, the efficiencies are somewhat higher than those achieved by the equivalent five variable designs for logistic regression. 

There are three main points of interest: (1) the $D$-efficiencies are higher for the five variable design due to the smaller number of variables in the maximal model leading to less imbalance in the variable values in the sub-designs; (2) for both the five-variable and 10-variable designs, the $D$-efficiency increases with the size of the model, reflecting the construction method of maximizing~\eqref{eq:PoissIC} for the maximal model; and (3) the spread of the $D$-efficiencies decreases as $\kappa$ increases, making the prior distribution more concentrated. In both examples and for all $\kappa$ values, the minimum efficiency is greater than $\sim$0.4, and the mean efficiency is greater than $\sim$0.55. For smaller $\kappa$ and models with larger numbers of variables, the designs often have much higher $D$-efficiencies.

For this example, we also assess the performance of minimum aberration fractional factorial designs of resolution V (with $N=16$ runs for five variables) and resolution IV (with $N=32$ runs for 10 variables), see Figure~\ref{fig:poissdefflocal}. Although these designs are $D$-optimal for the linear model, they perform uniformly poorly under log-linear regression, and much worse than the robust minimally-supported designs. Their efficiencies are particularly low for the larger values of $\kappa$, where the variance of the response is least constant. 

\subsection{Model selection results}\label{sec:poisresults}

Simulations to assess the performance of the designs for model selection were conducted as described in Section~\ref{sec:binresults} except that, in step (iii), the model parameters were estimated using maximum likelihood and a model was chosen using AIC. Figure~\ref{fig:poisssim} shows the results for five variable and 10 variable studies with $\kappa = 1$. Results in both cases are very encouraging, with almost uniformly high power and low type I error rates ($<0.2$). For data-generating models with only one active variable (models 1-5 for the five variable experiment and models 1-10 for the 10 variable experiment), the truly active variable is occasionally missed, and another variable is identified as active. These errors lead to slightly lower power for these models, and non-zero FDR. For models with larger numbers of variables, all active variables are successfully identified (power equal to 1). For the five variable study, the FDR is consistently just below 0.3, corresponding to a maximum of about one non-active variable being incorrectly identified as active. For the 10 variable study, no screening errors are made for models containing three active variables (model 56 onwards). For both studies, the somewhat counter-intuitive result that performance improves for true models containing more active variables is explained by the construction method of the design (see Section~\ref{sec:poisic}), which focusses on the model containing the maximum number of variables. 

For this example, the model selection performance of the two fractional factorial designs was very similar to that of the robust designs. Hence, the factorial designs would be effective for discrimination between the competing models but would provide poor estimation of the selected model.

\newcommand{\poisssimscale}{.33}
\begin{figure}
\begin{center}
\begin{tabular}{cc}
(a) & (b) \\[-2ex]
\includegraphics[scale=\poisssimscale]{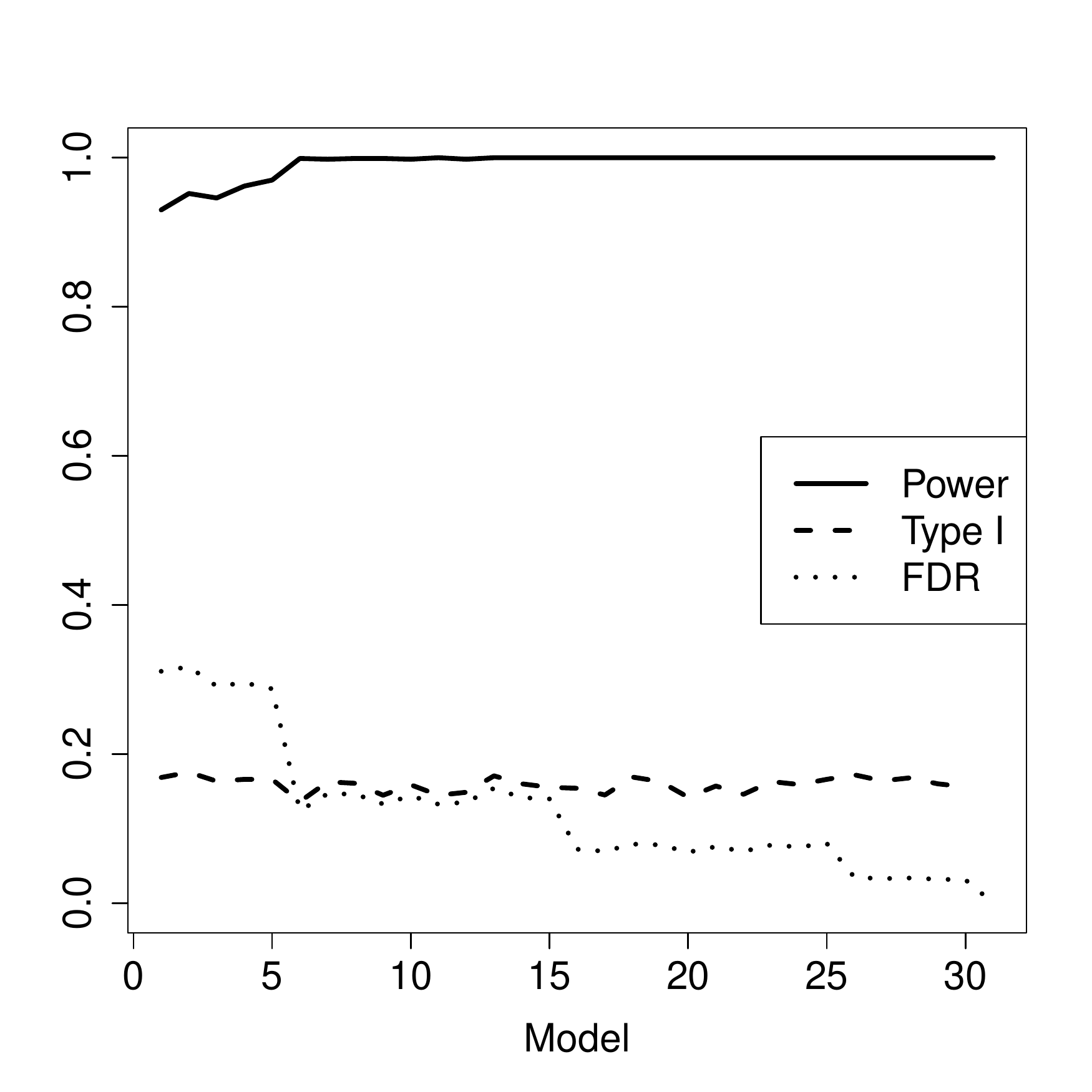} & \includegraphics[scale=\poisssimscale]{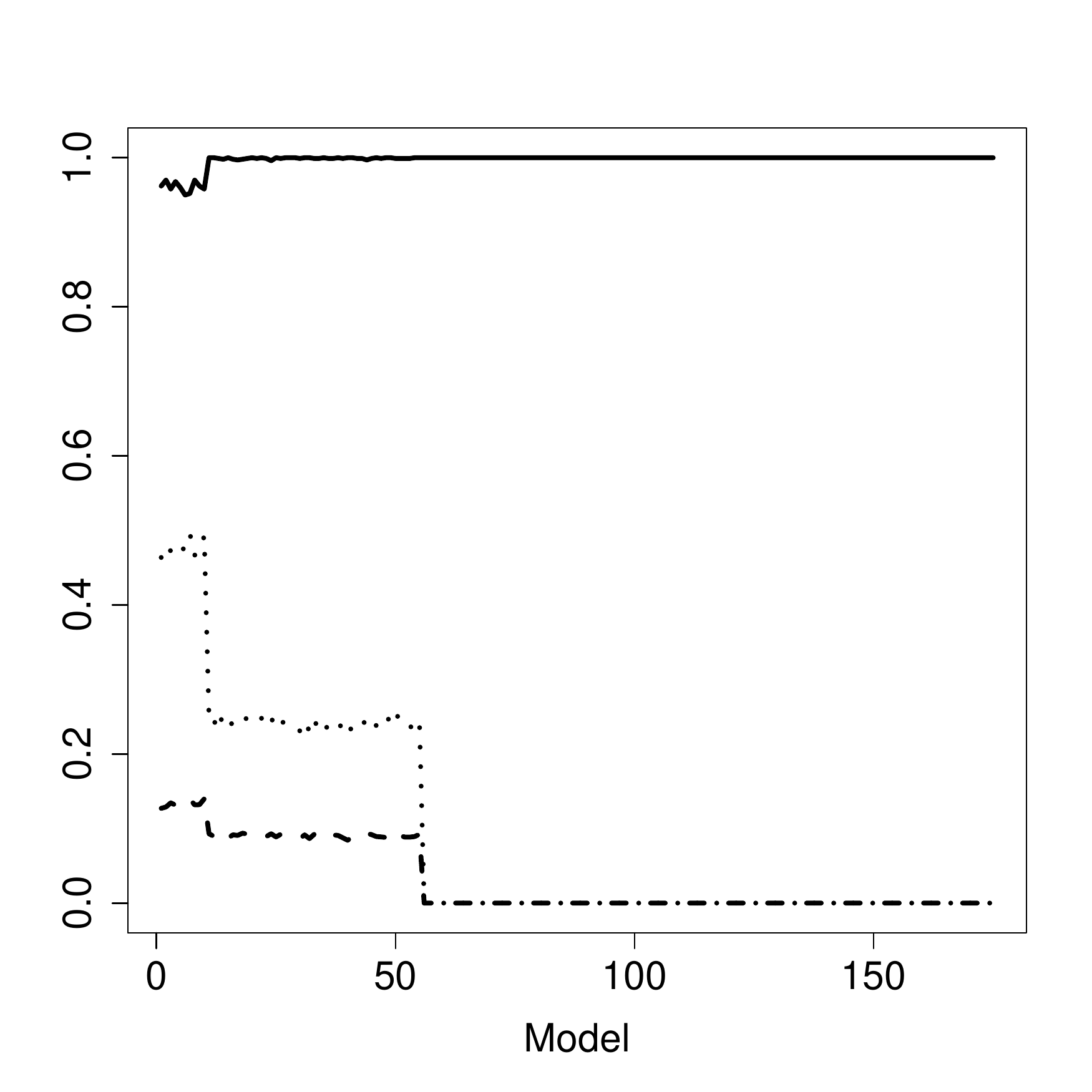}
\end{tabular}
\end{center}
\caption{\label{fig:poisssim}Power, type I error rate and FDR for log-linear regression for $\kappa=1$: (a) five variables and $N=16$; (b) 10 variables and $N=32$.}
\end{figure}

\section{Discussion and further research}\label{sec:disc}

This paper provides the first investigation of designs for screening variables under a generalised linear model. The results demonstrate that effective screening (high power with only moderate type I error rate and FDR) is achievable. For a binomial response and logistic regression, both design and model selection are more challenging than for a Poisson response, and larger designs are required to achieve good model selection results. For a binomial response, the results presented here can easily be extended to linear predictors that include products of variables representing interactions.  

Future work is needed to investigate in more detail the use of the GIC penalty with maximum penalised likelihood. In some experiments, it may also be necessary to choose the link function in addition to the linear predictor. Compromise designs for this situation were found by \citet{WLER2006}. In this paper, we have restricted the size of the model space under consideration by applying the principle of factor sparsity or by restricting the number of variables. For larger model spaces, the curse of dimensionality may prevent an all-subsets approach to model selection and alternative methods, such as sampling the model space \citep{SD2015} or shrinkage regression \citep{FHT2010}, would then need to be employed. Clearly, the choice of design, and any resultant ``confounding'' of model effects, will have an impact on any model selection procedure, including Bayesian and shrinkage methods. Investigations into the performance of these methods is another area for future work.

In the binomial and Poisson examples, we chose designs that ensured all models were estimable. This strategy is in contrast to the use of a design criterion tailored to model discrimination alone such as $T$-optimality \citep{AF1975a}, where the requirement of model estimability is often not met for nested models. An alternative approach is to generalise to multiple models those design selection methods that focus on both estimation and discrimination, such as the use of compound criteria \citep{Atkinson2008} or hybrid designs \citep{WWEL2008}. 

\section*{Acknowledgements}
D. C. Woods was supported by Fellowship EP/J018317/1 from the UK Engineering and Physical Sciences Research Council.

\section*{References}

\bibliography{biblio}

\end{document}